\documentstyle[aps,pre,amsfonts,graphicx]{revtex}

\newcommand{\beq}{\begin{equation}}
\newcommand{\eeq}{\end{equation}}
\def\erre{\hbox{\rm\rlap{I}\kern.1em R}}

\newcommand{\be}{\begin{equation}}

\begin{document}

\title{Hamiltonian dynamics and geometry of phase transitions in classical
XY models}

\author{Monica Cerruti-Sola$^{1,3,}$\cite{moni}, 
Cecilia Clementi$^{2,}$\cite{cecilia},
 and Marco Pettini$^{1,3,}$\cite{marco}}
\address{
$^1$Osservatorio Astrofisico di Arcetri, Largo E. Fermi 5, I-50125 Firenze,
 Italy \\
$^2$Department of Physics, University of California at San Diego, 
La Jolla, CA 92093-0319, USA \\
$^3$Istituto Nazionale per la Fisica della Materia, Unit\`a di Ricerca di
Firenze, Italy }

\date{\today}

\maketitle

\begin{abstract}
The Hamiltonian dynamics associated to classical, planar, Heisenberg XY 
models is investigated for two- and three-dimensional lattices.
Besides the conventional signatures of phase transitions, here obtained
through time averages of thermodynamical observables in place of ensemble 
averages, qualitatively new information is derived from the 
temperature dependence of Lyapunov exponents. A Riemannian geometrization
of newtonian dynamics suggests to consider other observables of geometric 
meaning tightly related with the largest Lyapunov exponent. The numerical
computation of these observables - unusual in the study of phase transitions -
sheds a new light on the microscopic dynamical counterpart of 
thermodynamics also pointing to the existence of some major change in the 
geometry of the mechanical manifolds at the thermodynamical transition. 
Through the microcanonical definition of the entropy, a relationship between
thermodynamics and the extrinsic geometry of the constant energy surfaces 
$\Sigma_E$ of phase space can be naturally established. In this framework, 
an approximate formula is worked out, determining a highly non-trivial 
relationship between temperature and topology of the $\Sigma_E$. Whence it can
be understood that the appearance of a phase transition must be tightly
related to a suitable major topology change of the $\Sigma_E$. This contributes
to the understanding of the origin of phase transitions in the
microcanonical ensemble.
\end{abstract}
\newpage

\section{Introduction}
\label{intro}
The present paper deals with the study of the microscopic Hamiltonian 
dynamical phenomenology associated to thermodynamical phase transitions.
This general subject is addressed in the special case of planar, classical
Heisenberg XY models in two and three spatial dimensions.
A preliminary presentation of some of the results and ideas contained in 
this paper has been already given in \cite{CCCP}.

There are several reasons to tackle the Hamiltonian dynamical counterpart of
phase transitions. On the one hand, we might wonder whether our knowledge
of the already wide variety of dynamical properties of Hamiltonian systems
can be furtherly enriched by considering the dynamical signatures, if any,
of phase transitions. On the other hand, it is {\it a-priori} conceivable
that also the theoretical investigation of the phase transition phenomena
could benefit of a direct investigation of the natural microscopic dynamics.
In fact, from a very general point of view, we can argue that in those times
where microscopic dynamics was completely unaccessible to any kind of 
investigation, statistical mechanics has been invented just to replace 
dynamics. During the last decades, the advent of powerful computers 
has made
possible, to some extent, a direct access to microscopic dynamics through the
so called molecular dynamical simulations of the statistical properties of
"macroscopic" systems.

Molecular dynamics can be either considered as a mere alternative to Monte
Carlo methods in practical computations, or it can be also seen as a possible
link to concepts and methods (those of nonlinear Hamiltonian dynamics) 
that could deepen our insight about phase transitions. 
In fact, by construction,
the ergodic invariant measure of the Monte Carlo stochastic dynamics, commonly
used in numerical statistical mechanics, is
the canonical Gibbs distribution, whereas there is no general result that 
guarantees the ergodicity and mixing of natural (Hamiltonian) dynamics. 
Thus, the general interest for any contribution that helps in clarifying 
under what conditions equilibrium statistical mechanics correctly describes
the average properties of a large collection of particles, safely replacing
their microscopic dynamical description. 

Actually, as it has been already shown and confirmed by the 
results reported below, there are some intrinsically dynamical observables
that clearly signal the existence of a phase transition.
Notably, Lyapunov exponents appear as sensitive measurements for phase
transitions.
They are also probes of a hidden geometry of the dynamics, because Lyapunov 
exponents depend on the geometry of certain ``mechanical manifolds''
whose geodesic flows coincide with the natural motions.
Therefore, a peculiar energy -- or temperature -- dependence of the largest 
Lyapunov exponent at a phase transition point also reflects some important
change in the geometry of the mechanical manifolds.

As we shall discuss throughout the present paper, also the topology of these
manifolds has been discovered to play a relevant role in the phase transition
phenomena (PTP).

Another strong reason of interest for the Hamiltonian dynamical
counterpart of PTP is related to the equivalence problem of statistical 
ensembles. Hamiltonian dynamics has its most natural and tight relationship
with microcanonical ensemble. Now, the well known equivalence among
all the statistical ensembles in the thermodynamic limit is valid in general 
in the absence of thermodynamic singularities, i.e. in the absence of phase 
transitions.  This is not a difficulty for statistical mechanics 
as it might seem at first sight
\cite{Gallavotti}, rather, this is a very interesting and intriguing point.

The inequivalence  of canonical and microcanonical ensembles in presence of 
a phase transition has been analytically shown for a particular model by 
Hertel and Thirring \cite{Thirring}, it is mainly revealed by the 
appearance
of negative values of the specific heat and has been discussed by several
authors \cite{Lyndenbell,Gross}.

The microcanonical description of phase transitions seems also to offer many
advantages in tackling first order phase transitions \cite{Gross2}, and 
seems considerably
less affected by finite-size scaling effects with respect to the canonical
ensemble description \cite{Gross1}.
This non-equivalence problem, together with certain advantages of the
microcanonical ensemble, strenghtens the interest for the Hamiltonian dynamical
counterpart of PTP. Let us briefly mention the existing contributions in the
field. 

Butera and Caravati \cite{Butera},
considering an XY model in two dimensions, found that 
the temperature dependence of the largest Lyapunov exponent changes just
near the critical temperature $T_c$ of the
Kosterlitz-Thouless phase transition.  Other interesting aspects of the
Hamiltonian dynamics of the XY model in two dimensions have been extensively
considered in \cite{Leoncini}, where a very rich phenomenology is reported.
Recently, the behaviour of Lyapunov exponents
has been studied in Hamiltonian dynamical systems: {\it i)} with long-range
interactions \cite{Rapisarda,Ruffo,Antoni}, {\it ii)} describing either 
clusters of particles or magnetic or gravitational models exhibiting phase 
transitions, {\it iii)} in classical
lattice field theories with $O(1)$, $O(2)$ and $O(4)$ global symmetries
in two and three space dimensions \cite{CCP1,CCCPPG}, {\it iv)} in the XY 
model in two and three space dimensions
\cite{CCCP}, {\it v)} in the "$\Theta$ - transition" of homopolymeric chains 
\cite{polimeri}.
The pattern of $\lambda(T)$ close to the critical temperature $T_c$ is
model-dependent. The behaviour of Lyapunov exponents near the transition point
has been considered also in the case of first- order phase transitions
\cite{Dellago,Mehra}. It is also worth mentioning the very intriguing result
of Ref.\cite{Berry}, where a glassy transition is accompanied by a sharp
jump of $\lambda(T)$. 

$\lambda(T)$ always detects a phase transition and, even if
its pattern close to the critical temperature $T_c$ is
model-dependent, it can be used as an order parameter -- of dynamical origin --
also in the absence of a standard order parameter (as in the case of the
mentioned "$\Theta$-transition" of homopolymers and of the glassy transition
in amorphous materials).
This appears of great prospective interest also in the light of recently
developed analytical methods to compute Lyapunov exponents (see Section IV). 

Among Hamiltonian models with long-range
interactions exhibiting phase
transitions, the most extensively studied is the
mean-field XY model \cite{Ruffo,Firpo,Ruffo_prl,Ruffo_talk}, whose
equilibrium statistical mechanics
is exactly described, in the thermodynamic limit, by mean-field
theory \cite{Ruffo}. In this system, 
the theoretically predicted temperature dependence of the largest Lyapunov 
exponent $\lambda$ displays a non-analytic behavior at the phase transition 
point. 

The aims of the present paper are
\begin{itemize}
\item 
to investigate the dynamical phenomenology of Kosterlitz-Thouless and of 
second order phase transitions in the $2d$ and $3d$ classical
Heisenberg XY models 
respectively;
\item
to highlight the microscopic dynamical counterpart of phase transitions
through the temperature dependence of the Lyapunov exponents, also providing
some physical interpretation of abstract quantities involved in
the geometric theory of chaos (in particular among vorticity, Lyapunov 
exponents and sectional curvatures of configuration space);
\item
to discuss the hypothesis that phase transition phenomena could be originated
by suitable changes in the topology of the  constant energy hypersurfaces of 
phase space, therefore hinting to a mathematical characterization of phase 
transitions in the microcanonical ensemble.
\end{itemize}

The paper is organized as follows: Sections $II$ and $III$ are devoted to 
the dynamical investigation of the $2d$ and $3d$ XY models respectively.
In Section $IV$ the geometric description of chaos is considered, with 
the analytic derivation of the temperature dependence of the largest Lyapunov 
exponent, the geometric signatures of a second-order phase transition and 
the topological hypothesis. 
Section $V$ contains a presentation of the relationship between the extrinsic
geometry and topology of the energy hypersurfaces of phase space and 
thermodynamics; the results of some numeric computations are also reported.
Finally, Section $VI$ is devoted to summarize the achievements reported in 
the present paper and to discuss their meaning.

\medskip
\section{$2d$ XY model}
\medskip

We considered a system of planar, classical ``spins'' (in fact rotators) 
on a square lattice of $N=n\times n$ sites, and interacting through
the ferromagnetic interaction $V=-\sum_{\langle i,j\rangle}J{\bf S}_{i}
\cdot {\bf S}_{j}$ (where $\vert{\bf S}_i\vert =1)$. 
The addition of standard, i.e. quadratic, kinetic energy
term leads to the following choice of the Hamiltonian
\beq
H= \sum_{i,j=1}^{n} \left \{ \frac{p_{i,j}^2}{2}+J[2- \cos(q_{i+1,j}-q_{i,j})
-\cos(q_{i,j+1}-q_{i,j})]\right\}~~,
\label{xy2d}
\eeq
where $q_{i,j}$ are the angles with respect to a fixed direction on
the reference plane of the system.
In the usual definition of the XY model both the kinetic term and the
constant term $2JN$ are lacking; however, their contribution does not modify
the thermodynamic averages (because they usually depend only on the 
configurational partition function, 
$Z_C=\int\prod_{i=1}^N dq_i\exp[-\beta V(q)]$, 
the momenta being trivially integrable when the kinetic energy is quadratic).
Thus, as we tackle classical systems, the choice of a quadratic
 kinetic energy term
is natural because it corresponds to  $\frac{1}{2}\sum_{i=1}^N \vert
{\bf\dot S}_i\vert^2$, written in terms of the momenta $p_{i,j}$ canonically
conjugated to the lagrangian coordinates $q_{i,j}$. The constant term $2JN$ is 
introduced to make the low energy expansion of Eq. (\ref{xy2d}) coincident
with the
Hamiltonian of a system of weakly coupled harmonic oscillators.

The theory predicts for this model a Kosterlitz-Thouless phase transition
occurring at a critical temperature
 estimated around $T_c\sim J$. Many Monte Carlo
simulations of this model have been done in order to check the predictions
of the theory. Among them, we quote those of
Tobochnik and Chester \cite{TobChes} and of Gupta and Baillie \cite{Gupta}
which, on the basis of accurate numerical analysis, confirmed the predictions 
of the theory and fixed the critical temperature at $T_c=0.89$ ($J=1$).

The analysis of the present work is based on the numerical integration
of the equations of motion derived from Hamiltonian (\ref{xy2d}).
The numerical integration is performed by means of
a bilateral, third order, symplectic algorithm \cite{Lapo}, and it is 
repeated at several values of the energy density
$\epsilon = E/N$ ($E$ is the total energy of the system which depends upon 
the choice of the initial conditions). 
While the Monte Carlo simulations perform statistical averages in the canonical
ensemble, Hamiltonian dynamics has its statistical counterpart in the 
microcanonical ensemble. Statistical averages are here replaced by time 
averages of relevant observables. In this perspective, from the microcanonical
definition of temperature $1/T=\partial S/
\partial E$, where $S$ is the entropy,  
two definitions of temperature are available: $T=\frac{2}{N}\langle K\rangle$
(where $K$ is the kinetic energy per degree of freedom), if 
$S=\log \int\prod_{i=1}^N dq_i dp_i \Theta (H(p,q) - E)$, where $\Theta
(\cdot)$ is the Heaviside step function, and
$\tilde T=\left[\left(\frac{N}{2}-1\right)\langle K^{-1}\rangle\right]^{-1}$,
if $S=\log \int\prod_{i=1}^N dq_i dp_i \delta (H(p,q) - E)$ \cite{Pearson}.
$T$ (or $\tilde T$) are numerically
determined by measuring the time average of the kinetic energy $K$ 
per degree of freedom (or its inverse), i.e. 
$T=\lim_{t\rightarrow\infty}\frac{2}{N}
\frac{1}{t}\int_0^td\tau K(\tau)$ (and similarly for $\tilde T$).
There is no appreciable difference in the outcomes of the computations of 
temperature according to these two definitions.

\medskip
\subsection{Dynamical analysis of thermodynamical observables}
\medskip

\subsubsection{Order parameter}

The order parameter for a system of planar ``spins'' whose Hamiltonian  
is invariant under the action of the group
$O(2)$, is the bidimensional vector
\beq
{\bf M}= (\sum_{i,j=1}^{n}{\bf S}^x_{i,j},
\sum_{i,j=1}^{n}{\bf S}^y_{i,j})\equiv (\sum_{i,j=1}^{n}\cos q_{i,j},
\sum_{i,j=1}^{n} \sin q_{i,j}),
\label{order_par}
\eeq
which describes the mean spin orientation field.
After the 
Mermin-Wagner theorem, we know that no symmetry-breaking transition can
occur in one and two dimensional systems with a continuous symmetry and
nearest-neighbour interactions. This means that, at any non-vanishing 
temperature,
the statistical average of the total magnetization vector is necessarily
zero in the thermodynamic limit.
However, a vanishing magnetization is not necessarily expected
when computed by means of Hamiltonian dynamics at
finite $N$.
In fact, statistical averages are equivalent to averages computed through
suitable markovian Monte Carlo dynamics that {\it a-priori} can reach
any region of phase space, 
whereas in principle a true ergodicity breaking is possible in
the case of differentiable dynamics. Also an "effective" ergodicity breaking
of differentiable dynamics is possible, when the relaxation times -- of time
to ensemble averages -- are very fastly increasing with $N$ \cite{Palmer}.

This model has two integrable limits: 
coupled harmonic oscillators and free rotators, 
at low and high temperatures respectively. Hereafter, $T$ is meant in units
of the coupling constant $J$.

\begin{figure}
\centerline{\includegraphics[width=0.80\linewidth]{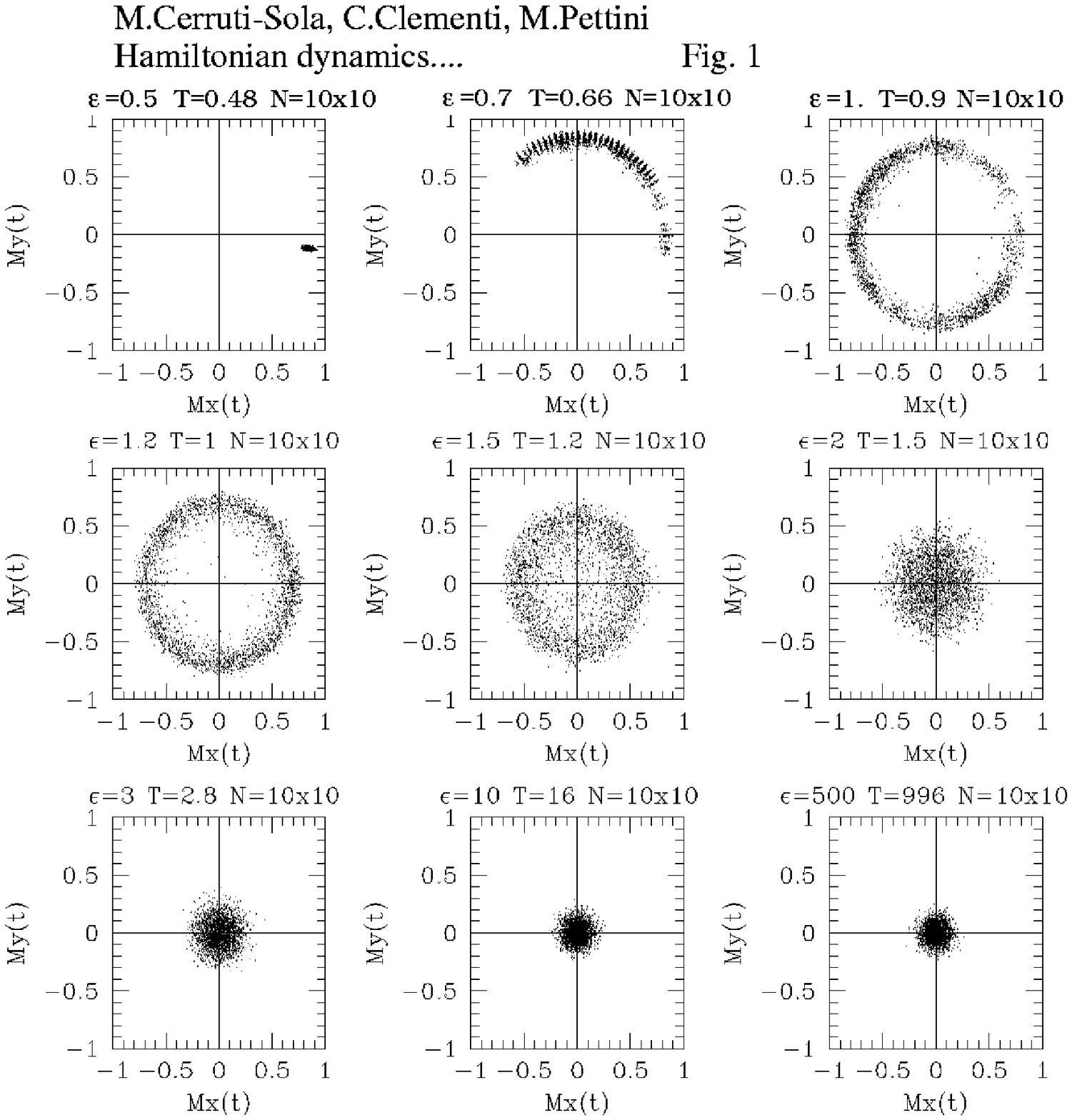}}
\caption{ The magnetization vector ${\bf M}(t)$ computed along a trajectory
for the $2d$ XY model at different temperatures on a lattice 
of $N= 10 \times 10$. Each point represents a vector ${\bf M}(t)$ at 
a certain time $t$. }
\label{figura.spin2d.10e10}
\end{figure}

For a lattice of $N = 10 \times10$ sites,
Figure \ref{figura.spin2d.10e10} shows that 
at low temperatures ($T<0.5$)-- being the system almost harmonic --
we can observe a persistent
memory of the total magnetization associated with the initial condition,
which, on the typical time scales of our numeric simulations ($10^6$ units 
of proper time), looks almost frozen.

By raising the temperature above a first threshold $T_0\simeq 0.6$,
the total magnetization vector -- observed on the same time scale -- 
starts rotating on the plane where it is confined.
A further increase of the temperature induces a faster rotation of the
magnetization vector together with a slight reduction of its average modulus.

At temperatures slightly greater than $1$, we observe that already at 
$N=10 \times10$ a random variation of the direction and of the modulus of
the vector ${\bf M}(t)$ sets in. 

At $T>1.2$, we observe a fast relaxation and, at high temperatures 
($T\simeq 10$), a sort
of saturation of chaos.

At a first glance, the results reported in Fig. 
\ref{figura.spin2d.10e10} could suggest 
the presence of a phase transition associated with the breaking of the
$O(2)$ symmetry.
In fact, having in mind the Landau theory, the ring-shaped distribution of
the instantaneous magnetization shown
by Fig. \ref{figura.spin2d.10e10} is the typical signature of an
$O(2)$-broken symmetry phase and the spot-like patterns around zero are
proper to the unbroken symmetry phase.

\begin{figure}
\centerline{\includegraphics[width=0.80\linewidth]{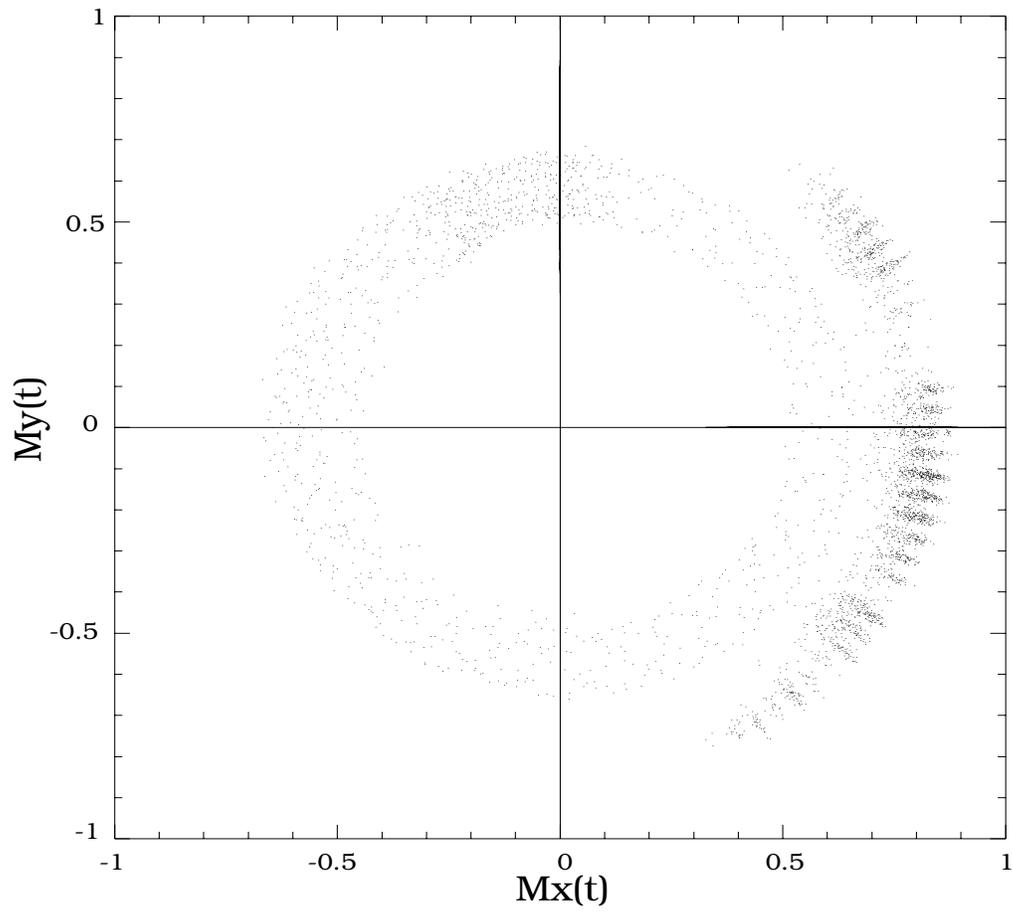}}
\caption{ The magnetization vector ${\bf M}(t)$ at the temperature $T=0.74$,
corresponding to the specific energy $\epsilon = 0.8$ and computed in 
a time interval $\Delta t = 10^5$, with a random initial configuration,
on lattices of $N = 10 \times 10$ (external points)
 and of $N = 200 \times 200$ 
(internal points). }
\label{fig.spin2d.t0.74}
\end{figure}

The apparent contradiction of these results with the Mermin-Wagner theorem
is resolved by checking whether the observed phenomenology is stable with
$N$. Thus, some simulations have been performed at larger values of $N$. 
At any temperature, we found that the average modulus 
$\langle\vert{\bf M}(t)\vert\rangle_t$ of the vector ${\bf M}(t)$, 
computed along the trajectory, systematically
decreases by increasing $N$. However, 
for temperatures lower than $T_0$, the $N$-dependence 
of the order parameter is very weak, whereas,
for temperatures greater than $T_0$, the $N$-dependence 
of the order parameter is rather strong.
In Fig. \ref{fig.spin2d.t0.74} two extreme cases
($N=10 \times10$ and $N=200 \times200$) are shown for $T = 0.74$.
The systematic trend of $\langle\vert{\bf M}(t)\vert\rangle$ 
toward smaller values at increasing $N$ is 
consistent with its expected vanishing in the limit $N\rightarrow\infty$.

\begin{figure}
\centerline{\includegraphics[width=0.80\linewidth]{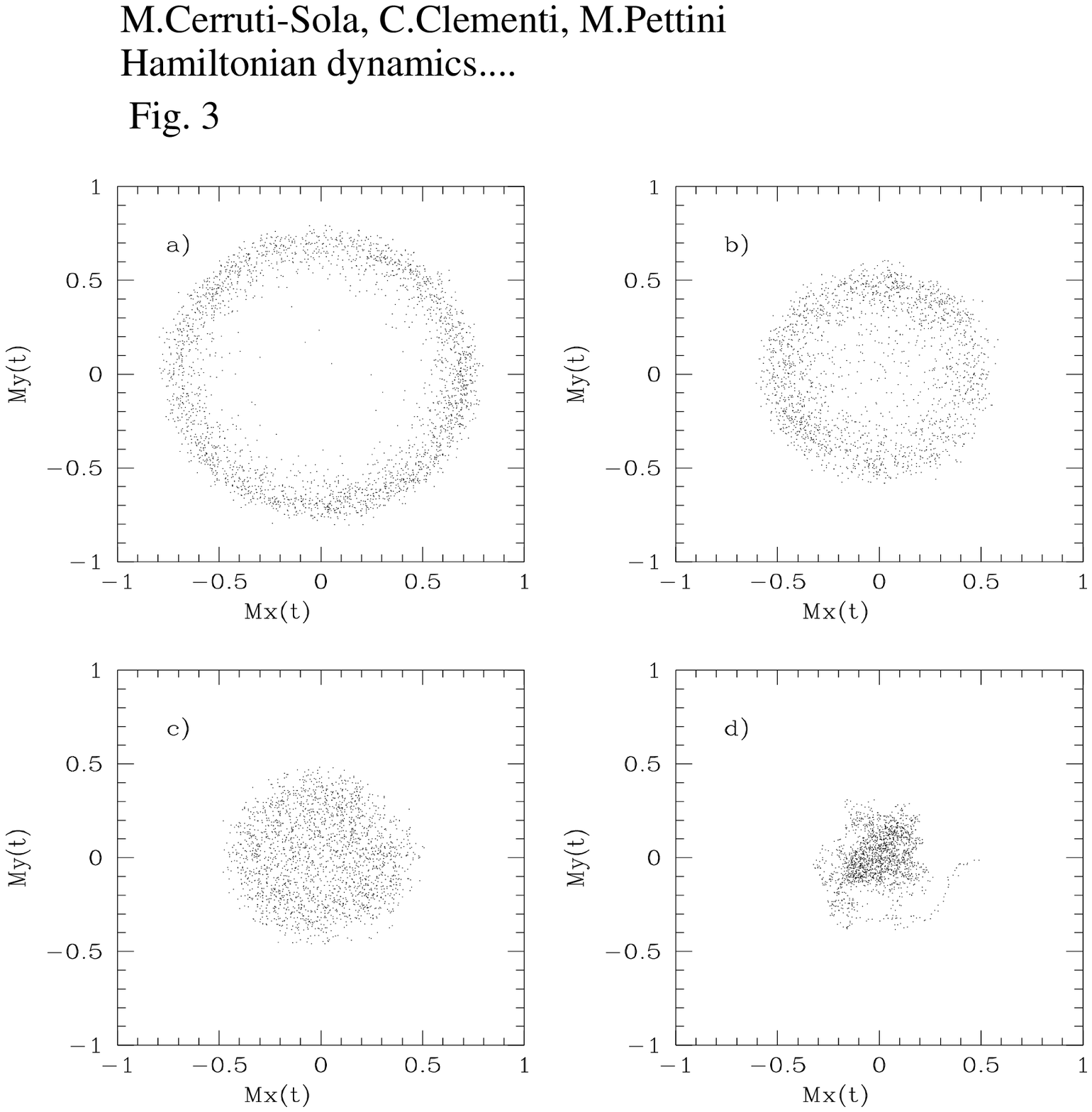}}
\caption{ The magnetization vector ${\bf M}(t)$ at the temperature $T=1$,
corresponding to the energy $\epsilon = 1.2$, computed in
a time interval $\Delta t = 10^5$, with a random initial configuration on 
lattices of {\it a)} $N = 10 \times 10$, {\it b)} $N = 50 \times 50$,
{\it c)} $N = 100 \times 100$ and {\it d)} $N = 200 \times 200$ sites, 
respectively. }
\label{fig.spin2d.t1}
\end{figure}

At $T = 1$, Fig. \ref{fig.spin2d.t1} shows that,
when the lattice dimension is greater than $50\times 50$,
${\bf M}(t)$ displays random variations both in direction (in the interval
[0,$2\pi$]) and in 
modulus (between zero and a value which is smaller at larger  
$N$).

\medskip
\subsubsection{Specific heat}
\medskip
By means of the recasting of a standard formula which relates the average 
fluctuations of a generic observable computed in canonical and 
microcanonical ensembles \cite{LPV}, and by specializing it to the
kinetic energy fluctuations, one obtains a microcanonical estimate 
of the canonical specific heat
\begin{equation}
c_{V}(T)=\frac{C_V}{N}\rightarrow 
\cases{
c_{V}(\epsilon)=\displaystyle{ \frac{k_{B}d}{2}\left[1-\frac{N
d}{2}\frac{\langle K^{2}\rangle -\langle K\rangle
^{2}}{\langle K\rangle ^{2}}\right]^{-1}} ~~,\cr
T=T(\epsilon)\cr }
\label{specheat}
\end{equation}
where $d$ is the number of degrees of freedom for each particle.
Time averages of the kinetic energy fluctuations computed at any
given value of the energy density $\epsilon$ yield $C_V(T)$, according
to its parametric definition in Eq.(\ref{specheat}).

>From the microcanonical definition $1/C_V=\partial T(E)/\partial E$ of the
constant volume specific heat, a formula can be worked out \cite{Pearson},
which is exact at 
{\it any} value of $N$ (at variance with the expression (\ref{specheat})).
 It reads
\begin{equation}
c_V=\frac{C_V}{N}=[N - (N - 2)\langle K\rangle\langle K^{-1}\rangle]^{-1}
\label{cvmicro}
\end{equation}
and it is the natural expression to be used in Hamiltonian
dynamical simulations of
finite systems.

\begin{figure}
\centerline{\includegraphics[width=0.80\linewidth]{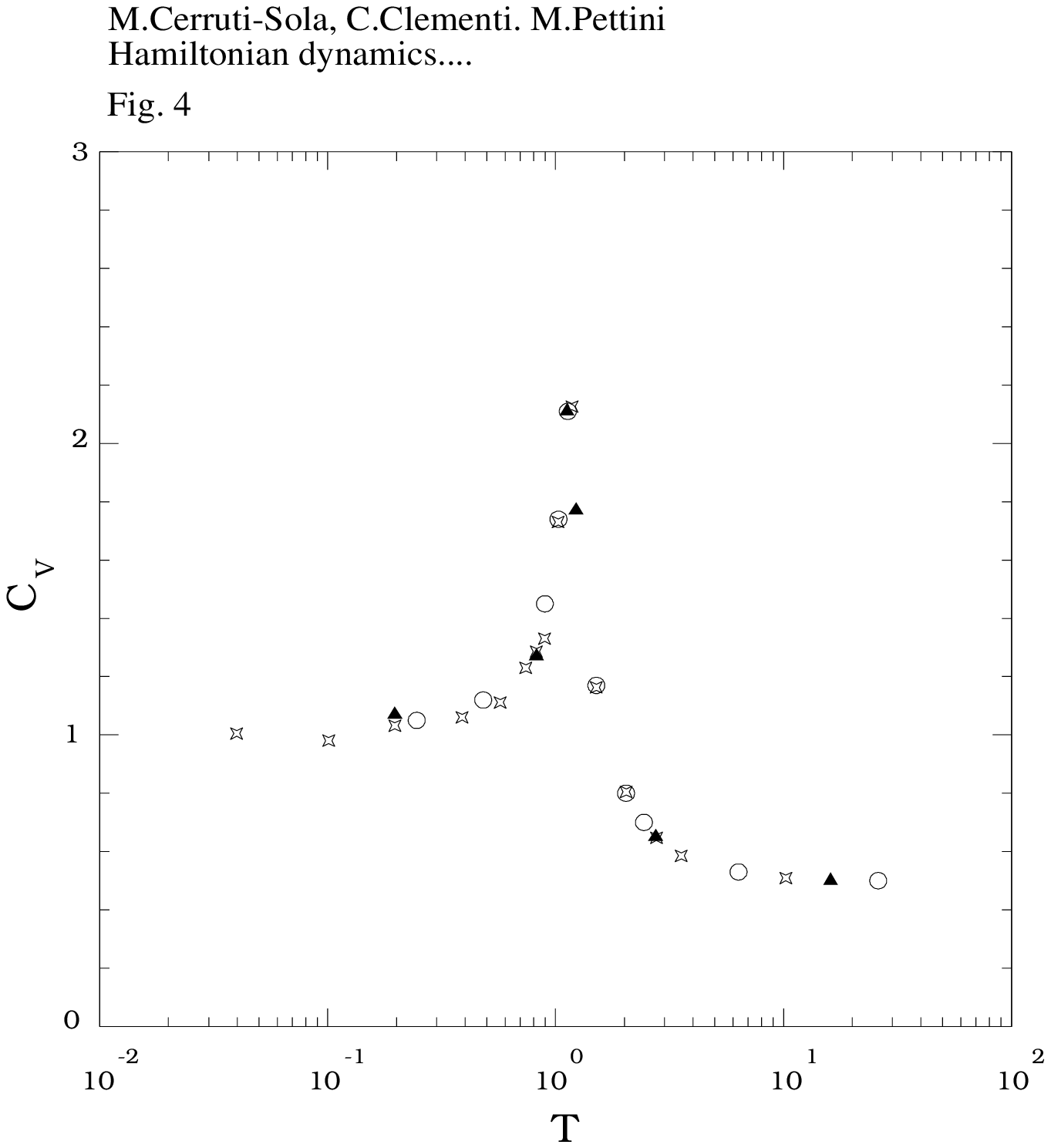}}
\caption{ Specific heat at constant volume 
computed by means of Eq. (\ref{cvmicro})
on a lattice of $N= 10 \times 10$ (open circles) and of $N= 15\times 15$
(full triangles). Starlike squares refer to specific heat values
computed by means of Eq. (\ref{specheat}) on a lattice of $N= 10 \times 10$
. }
\label{calspec_2d}
\end{figure}

The numerical simulations of the Hamiltonian dynamics of the $2d$
XY model -- computed with both Eqs.(\ref{specheat}) and (\ref{cvmicro}) -- 
yield a cuspy pattern for $c_V(T)$ peaked at $T\simeq 1$
(Fig. \ref{calspec_2d}).
This is in good agreement with the outcomes of canonical Monte Carlo 
simulations reported in Ref. \cite{TobChes,Gupta},
where a pronounced peak of $c_{V}(T)$ was detected at $T \simeq 1.02$.

By varying the lattice dimensions, the peak height remains
constant, in agreement with the absence of a symmetry-breaking phase
transition.  
\medskip
\subsubsection{Vorticity}
\medskip
Another thermodynamic observable which can be studied 
is the vorticity of the system. Let us briefly recall that if
the angular differences of nearby ``spins'' are small, we can suppose the
existence of a continuum limit function $\theta({\bf r})$ that conveniently
fits a given spatial configuration of the system.
Spin waves correspond to regular patterns of $\theta({\bf r})$, whereas the
appearance of a singularity in $\theta({\bf r})$ corresponds to a topological 
defect, or a vortex, in the ``spin'' configuration. When such a defect is 
present, along any closed path ${\cal C}$ that contains the centre of the 
defect, one has
\begin{equation}
\oint_{\cal C}\nabla\theta({\bf r})\cdot d{\bf r}= 2\pi q~,~~~~q=0,\pm 1,\pm 2,
\dots
\end{equation}
indicating the presence of a vortex whose intensity is $q$. For 
a lattice model with periodic boundary conditions, there is an equal number 
of vortices and antivortices (i.e. vortices rotating in opposite directions).
Thus, the vorticity of our model can be defined as the mean total number of 
equal sign vortices per unit volume.
In order to compute the vorticity ${\cal V}$ as a function of temperature,
we have averaged the number of positive vortices along the numerical phase 
space trajectories. On the lattice, ${\bf r}$ is replaced by the multi-index 
${\bf i}$ and $\nabla_\mu\theta_i= q_{{\bf i}+\mu} - q_{\bf i}$, then the
number of elementary vortices is counted: the discretized version of
$\oint_\square\nabla\theta\cdot d{\bf r}=1$ amounts to one elementary vortex 
on a plaquette. Thus ${\cal V}$ is obtained by summing over all the plaquettes.

Our results are in agreement with the values obtained 
by Tobochnik and Chester \cite{TobChes}
by means of Monte Carlo simulations with $N=60\times 60$. 

\begin{figure}
\centerline{\includegraphics[width=0.80\linewidth]{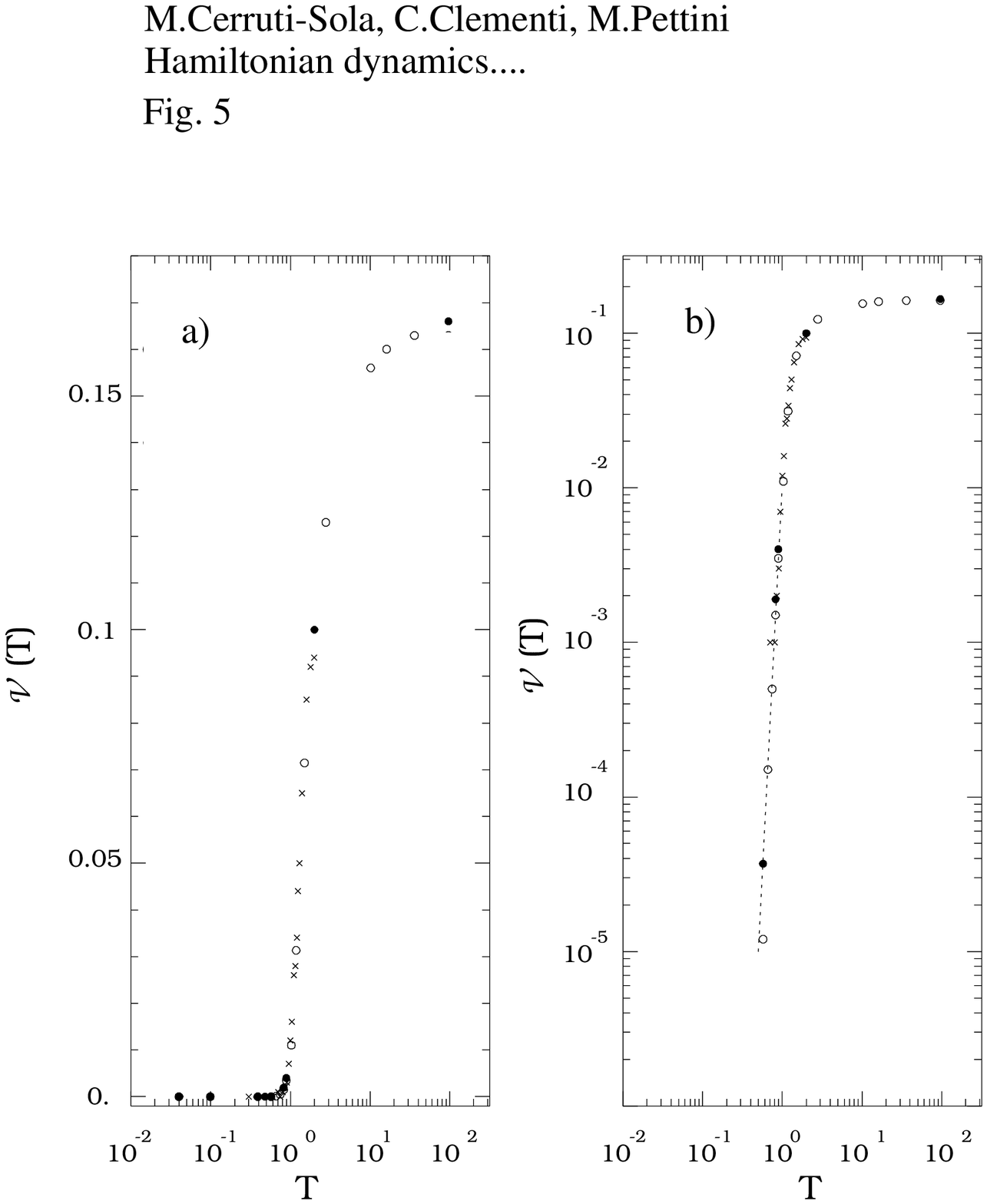}}
\caption{ Vorticity function (plotted in {\it a)} linear scale
and {\it b)} logarithmic scale) computed at different temperatures
for lattices of $N=10\times 10$ (open circles) and $N=40\times 40$
(full circles). The results of the Monte Carlo
simulations for a lattice of $N=60\times 60$ (crosses)
 are from \protect\cite{TobChes}. 
The dashed line represents the power
law ${\cal V}(t) \sim T^{10}$. }
\label{fig.vort2d}
\end{figure}

As shown in Fig. \ref{fig.vort2d},
on the $10\times 10$ lattice, 
the first vortex shows up at $T\sim0.6$
and on the $40\times 40$ lattice 
at $T\sim0.5$, when the
system changes its dynamical behavior,
increasing its chaoticity (see next Subsection).
 At lower temperatures, vortices are less probable,
due to the fact
that the formation of  vortex has a minimum energy cost.
Below $T\sim1$, the vortex density steeply grows 
with a power law ${\cal V}(T)\sim T^{10}$.
 The growth of ${\cal V}$ then slows down, until the saturation is reached at 
$T\sim10$.

\medskip
\subsection{Lyapunov exponents and chaoticity}
\medskip

\begin{figure}
\centerline{\includegraphics[width=0.80\linewidth]{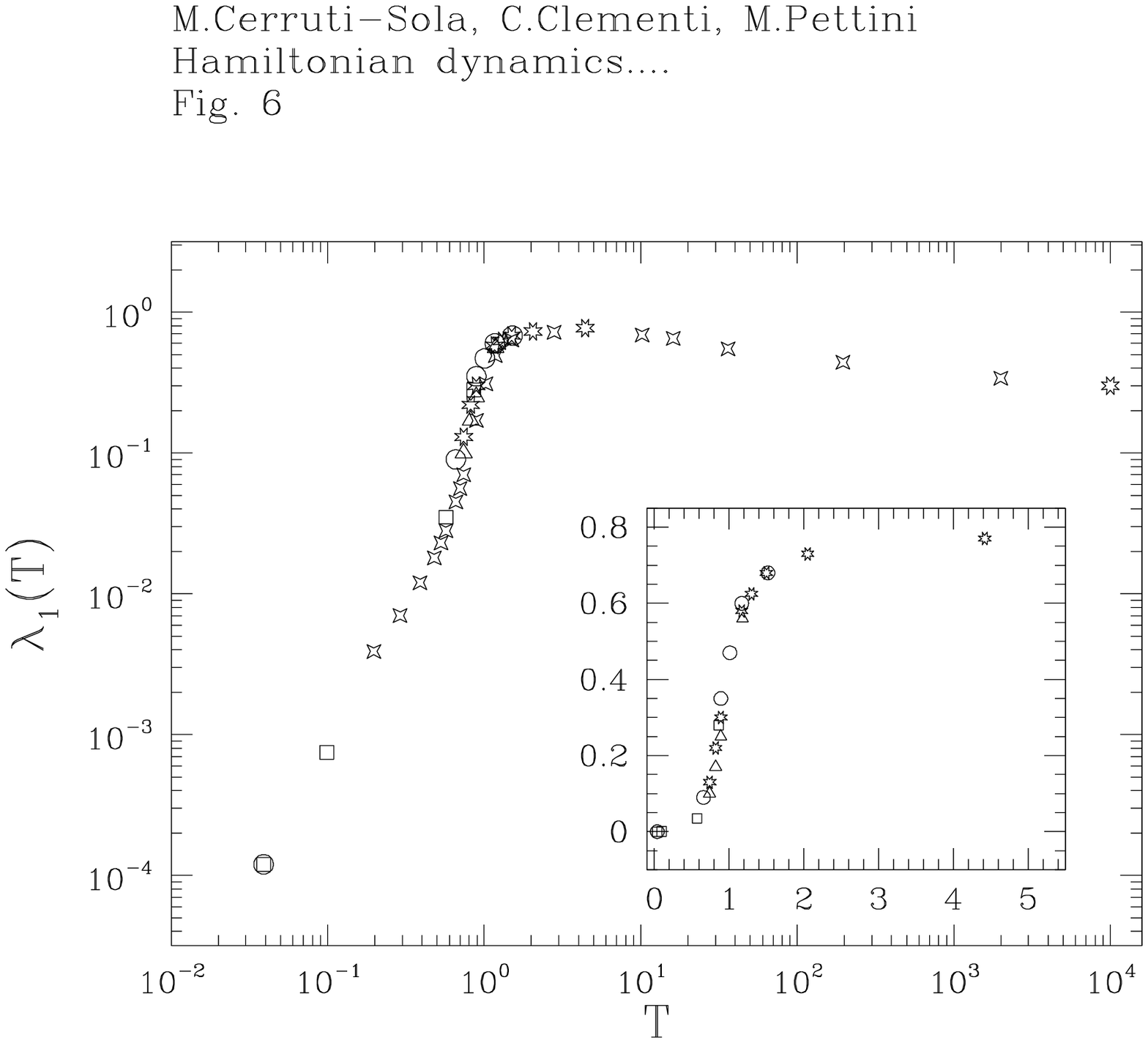}}
\caption{ The largest Lyapunov exponents computed on different lattice 
sizes:
$N = 10 \times 10$ (starred squares), $N= 20 \times 20$ (open triangles),
$N= 40 \times 40$ (open stars), $N= 50 \times 50$ (open squares) and 
$N = 100 \times 100$ (open circles). In the inset, symbols have the same
meaning. } 
\label{xy2d.lyap.num.fig}
\end{figure}

The values of the largest Lyapunov exponent $\lambda_{1}$
have been computed using the standard tangent dynamics equations [see Eqs.
(\ref{eqdintang}) and (\ref{bgs})], and are reported  
in Fig. \ref{xy2d.lyap.num.fig}.

Below $T\simeq 0.6$, 
the dynamical behavior is nearly the same as that of harmonic
oscillators and the excitations of the system are only ``spin-waves''.

In the interval $[0., 0.6]$, the observed temperature dependence
$\lambda_1(T) \sim T^2$ is equivalent to the 
$\lambda_1(\epsilon) \sim\epsilon^2$ dependence (since at low temperature  
$T(\epsilon) \propto \epsilon$), already found -- analytically and
 numerically --
in the quasi-harmonic regime of other systems and characteristic of weakly
chaotic dynamics \cite{CCP}.
 
Above $T \simeq0.6$, vortices begin to form and correspondingly the largest
Lyapunov exponent signals a "qualitative" change of the dynamics through a
steeper increase vs. $T$.

At $T \simeq0.9$, where the theory predicts a Kosterlitz - Thouless phase 
transition, $\lambda_1(T)$ displays an inflection point.

Finally, at high temperatures, the power law $\lambda_{1}(T)\sim T^{-1/6}$
is found.

\medskip
\section{$3d$ XY model}
\medskip

In order to extend the dynamical investigation 
to the case of second-order phase transitions, we have
studied a system described by an Hamiltonian having at the same time
the main characteristics
of the $2d$ model and the differences necessary to the appearance
of a spontaneous symmetry-breaking below a certain critical temperature.
 The model
we have chosen is such that the spin rotation is constrained on a plane and
only the lattice dimension has been increased, in order to elude the
``no go'' conditions of the Mermin-Wagner theorem.
This is simply achieved by tackling a system
defined on a cubic lattice of $N=n\times n\times n$ sites
and described by the Hamiltonian
\begin{eqnarray}
H&=& \sum_{i,j,k=1}^{n}
\{ \frac{p_{i,j,k}^{2}}{2}+J[3-\cos(q_{i+1,j,k}-q_{i,j,k})-\nonumber\\
&-&\cos(q_{i,j+1,k}-q_{i,j,k})-\cos(q_{i,j,k+1}-q_{i,j,k})] \}~~.
\end{eqnarray}

\medskip
\subsection{Dynamical analysis of thermodynamical observables}
\medskip
The basic thermodynamical phenomenology of a second-order phase transition is 
characterized 
by the existence of equilibrium
configurations that make the order parameter bifurcating away from zero
at some critical temperature $T_{c}$  and by a divergence of 
the specific heat $c_{V}(T)$ at
the same $T_{c}$. 
Therefore, this is the obvious starting point for the Hamiltonian dynamical
approach.

\medskip
\subsubsection{Order parameter}
\medskip
Below a critical value of the temperature,
the symmetry-breaking in a system invariant under the action of the
$O(2)$ group, appears
as the selection
 -- by the average magnetization vector of Eq. (\ref{order_par})--
 of a preferred direction  
 among all the possible, energetically equivalent choices. 
By increasing the lattice dimension, the symmetry breaking is therefore
characterized by a sort of simultaneous "freezing"
 of the direction of the order parameter ${\bf M}$ and of the
convergence of its modulus to a non-zero value.

\begin{figure}
\centerline{\includegraphics[width=0.80\linewidth]{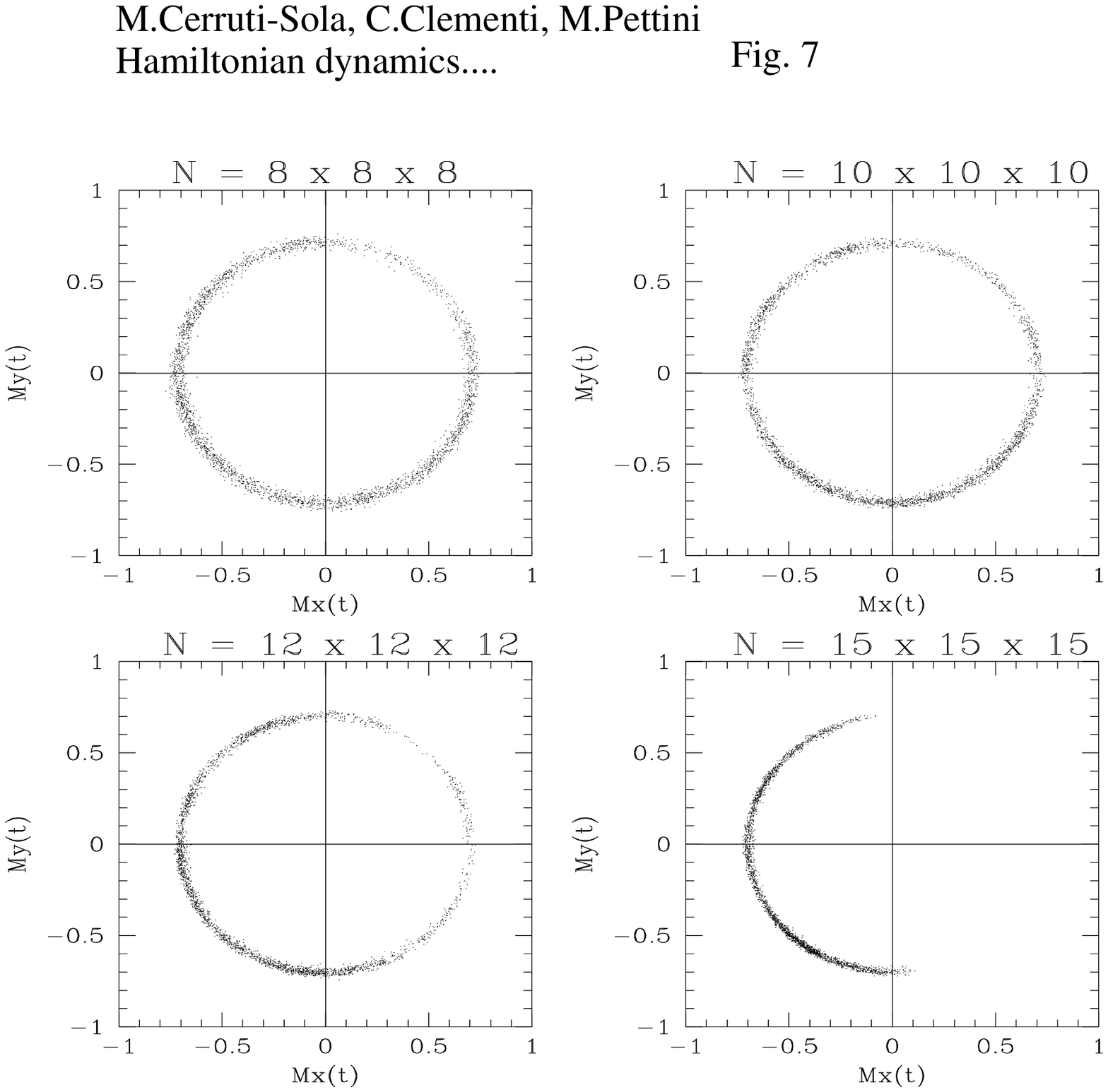}}
\caption{ The magnetization vector ${\bf M}(t)$, computed at the temperature
$T = 1.7$, on lattices of different sizes. By increasing the lattice 
dimensions,
the longitudinal fluctuations decrease. The time interval $\Delta t = 3.5
\times 10^4 - 8 \times 10^4$ is the same for the four simulations. }
\label{mag3d.e2}
\end{figure}

Figure \ref{mag3d.e2} shows that in the $3d$ lattice, 
at $T < 2$, i.e. in the broken-symmetry phase (as we shall see in the 
following),  
the dynamical simulations yield a thinner spread of the longitudinal
fluctuations by increasing $N$ -- that is, $| {\bf M} |$
oscillates by exhibiting a trend
to  converge to a
non-zero value --  and that the transverse 
fluctuations damp, ``fixing'' the direction of the oscillations.
This direction depends on the initial conditions.

\begin{figure}
\centerline{\includegraphics[width=0.80\linewidth]{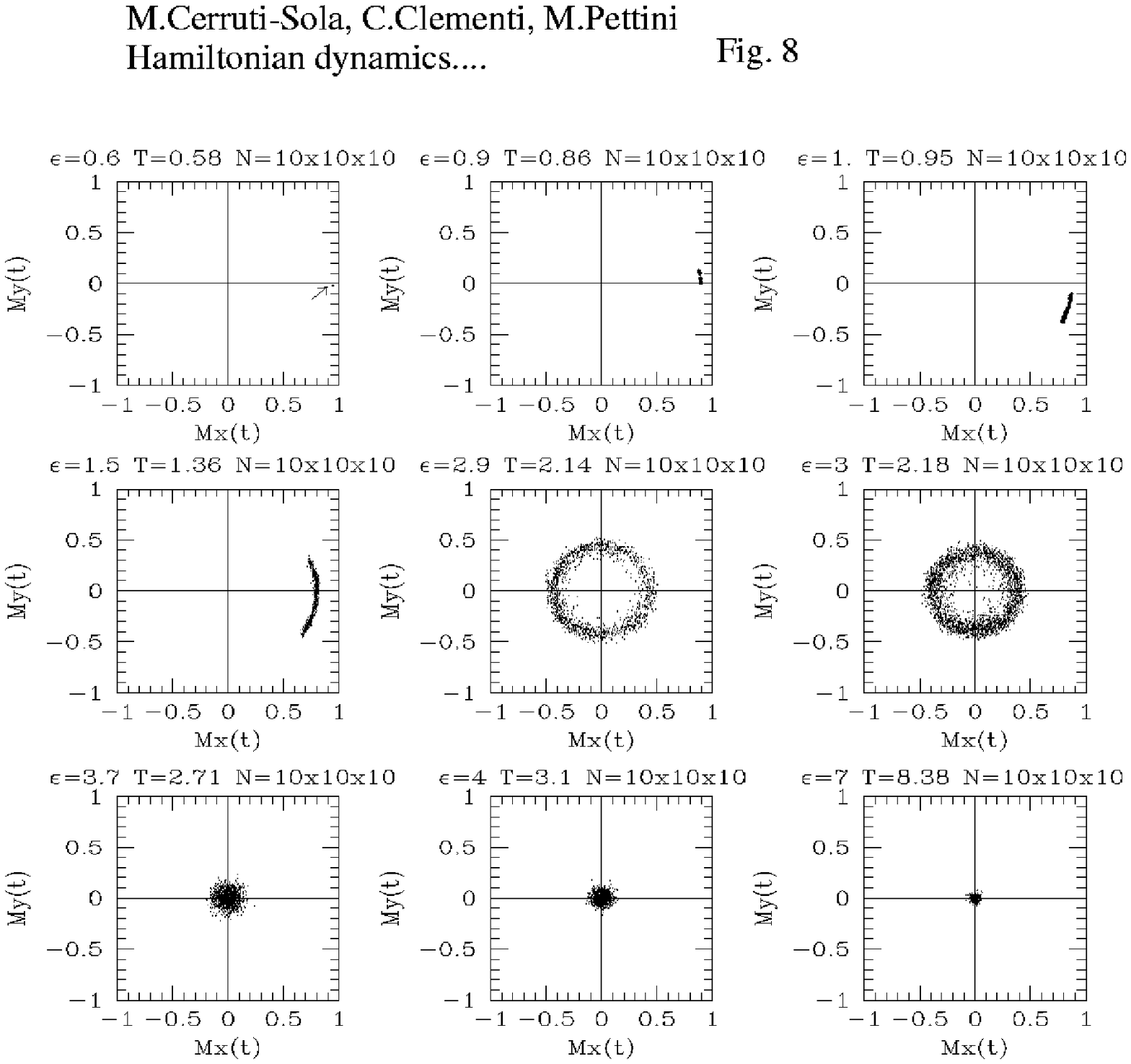}}
\caption{ The magnetization vector ${\bf M}(t)$ computed at different 
temperatures on a lattice of $N = 10 \times 10 \times 10$ spins. }
\label{fig.1.spin3d.9}
\end{figure}

Moreover, the dynamical analysis provides us with a better detail than a 
simple distinction between regular and chaotic dynamics. 
In fact, it is possible to 
distinguish between three different dynamical regimes 
(Fig. \ref{fig.1.spin3d.9}).

At low temperatures, up to $T \simeq 0.8$, one observes the
persistency of the initial direction and of an equilibrium value of the 
modulus $| {\bf M} |$ close to one.

At $0.8 < T < 2.2$, one observes transverse
oscillations, whose amplitude increases with temperature.

\begin{figure}
\centerline{\includegraphics[width=0.50\linewidth]{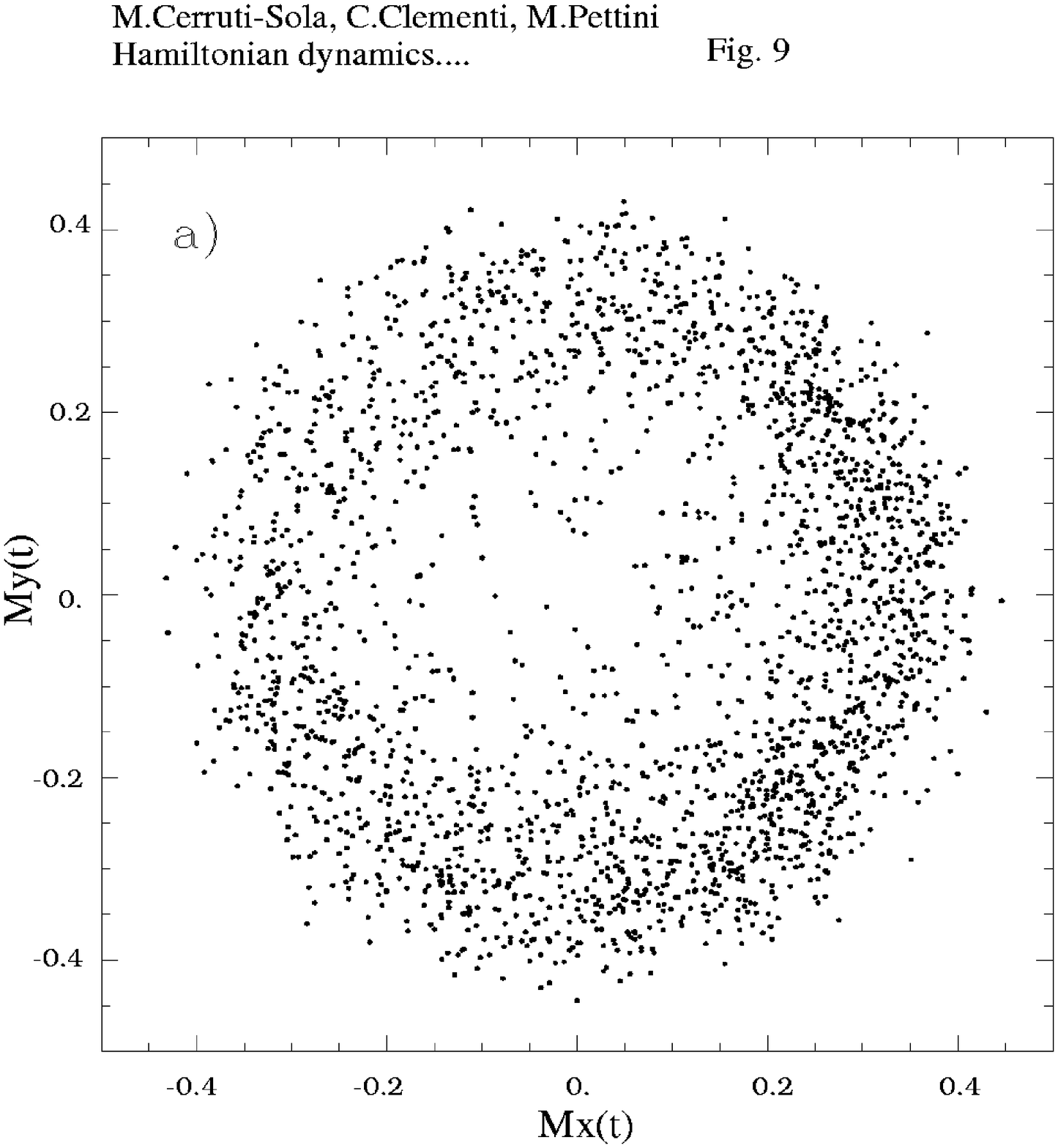}}
\medskip
\centerline{\includegraphics[width=0.50\linewidth]{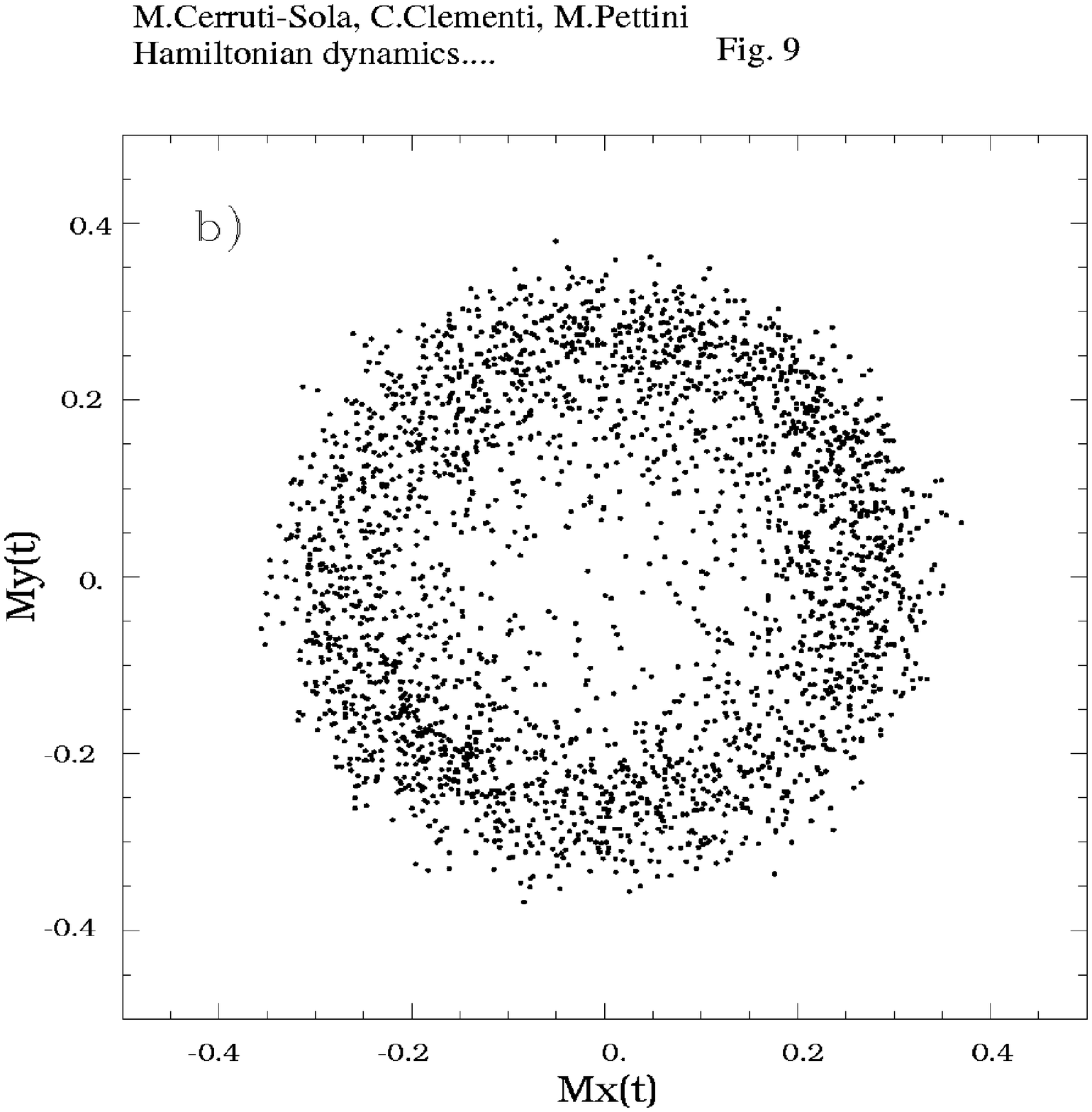}}
\caption{ The magnetization vector ${\bf M}(t)$ computed at 
the temperature $T = 2.22$ (slightly higher than the critical value)
on lattices of {\it a)} $N = 10 \times 10 \times 10$ 
and {\it b)}$N = 15 \times 15 \times 15$, respectively.
The time interval $\Delta t = 0.5 \times 10^4 - 1.5 \times 10^4$ is the same
for both simulations. } 
\label{3d.fig.spin.altat}
\end{figure}

At $T > 2.2$, 
the order parameter exhibits the features typical of an unbroken symmetry
phase.
In fact, it displays fluctuations peaked at zero, whose
dispersion decreases by increasing the temperature (bottom of
Fig. \ref{fig.1.spin3d.9}) and,
at a given temperature, by increasing the lattice volume (Fig.
\ref{3d.fig.spin.altat}a,b). 

\begin{figure}
\centerline{\includegraphics[width=0.80\linewidth]{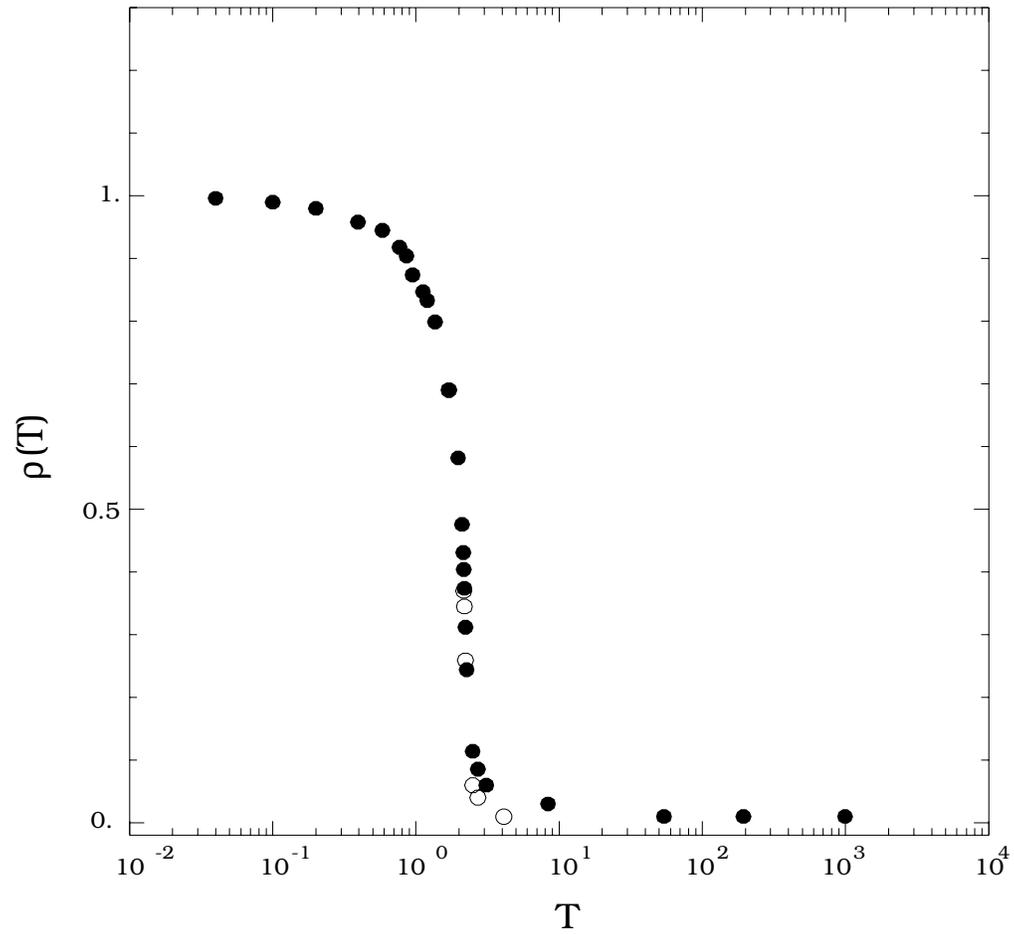}}
\caption{ The dynamical order parameter,
defined as the average of the modulus $| {\bf M}(t) |$ along a trajectory,
computed on lattices of
$N = 10 \times 10 \times 10$ (full circles)
and $N = 15 \times 15 \times 15$ (open circles). }
\label{parord.3d.fig}
\end{figure}

We can give an estimate of the order parameter by evaluating the average 
of the modulus $\langle | {\bf M}(t)|\rangle
= \rho(T)$. At $T < 2.2$, the $N$-dependence is given mainly by   
the rotation of the vector, while the longitudinal oscillations are moderate,
as shown in Fig. \ref{parord.3d.fig}. At temperatures above $T \simeq2.2$, 
we observe the squeezing of $\rho(T)$ to a small value.

The existence of a second order phase transition can be recognized
by comparing the temperature behavior and the $N$-dependence of 
the thermodynamic observables computed for
the $2d$ and the $3d$ models.
Both systems exhibit
the rotation of the magnetization
vector and small fluctuations of its modulus when they are considered on small
lattices.
In the $2d$ model the average modulus of the order parameter is theoretically 
expected to vanish logarithmically with $N$, what seems qualitatively 
compatible with the weak $N$ dependence shown in
Fig. \ref{fig.spin2d.t0.74}, whereas in the $3d$ model we observe a stability
with $N$ of $\langle| {\bf M} |\rangle$, suggesting the convergence
to a non-zero value of the order parameter also in the limit
$N\rightarrow\infty$, as shown in Fig. \ref{mag3d.e2}.

$T \simeq2.2$ is an approximate value of the critical temperature $T_c$ of the
second-order phase transition. This value will be refined in the following
Subsection. No finite-size scaling analysis has been performed for two
different reasons: {\it i)} our main concern is a qualitative phenomenological
analysis of the Hamiltonian dynamics of phase transitions rather than a 
very accurate quantitative analysis, {\it ii)} finite-size effects are much 
weaker in the microcanonical ensemble than in the canonical ensemble 
\cite{Gross1}.

\medskip
\subsubsection{Specific heat}
\begin{figure}
\centerline{\includegraphics[width=0.80\linewidth]{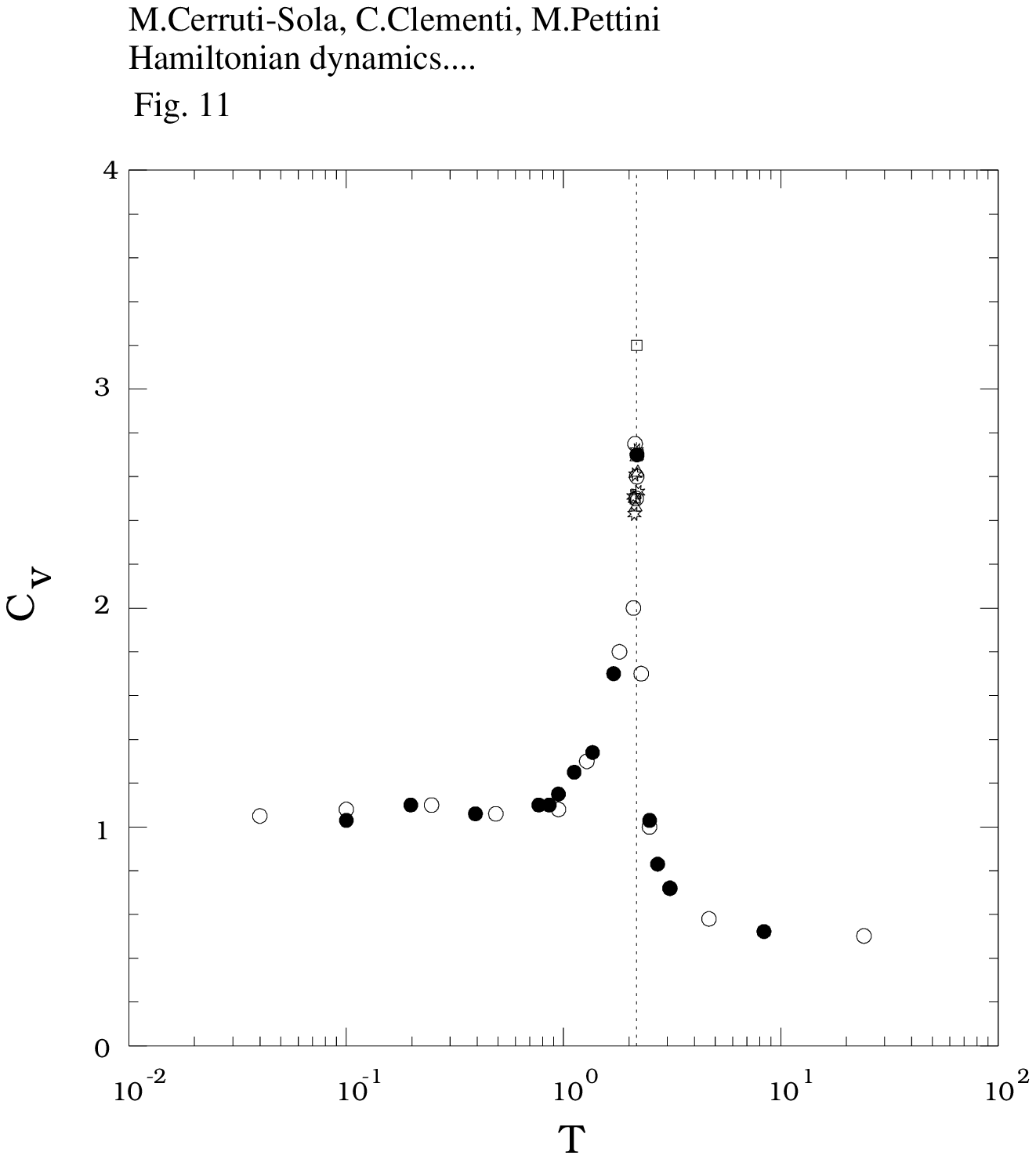}}
\caption{ Specific heat at constant volume for the $3d$ model, computed
by means of Eq. (\ref{cvmicro}) on
lattices of $N= 8 \times 8 \times 8$ (open triangles),
$N = 10 \times 10 \times 10$ (open circles), $N = 12 \times 12 \times 12$
(open stars) and $N= 15 \times 15 \times 15$ (open squares) .
Full circles refer to specific heat values computed by means of Eq. (\ref
{specheat}) on a lattice of $N= 10 \times 10 \times 10$. 
The dashed line points out the critical temperature $T_c \simeq 2.17$ at which
the phase transition occurs. }
\label{calspec.3d.fig}
\end{figure}

\medskip
As in the $2d$ model, numerical simulations of the Hamiltonian dynamics
have been performed with both Eqs.(\ref{specheat}) and (\ref{cvmicro}).
The outcomes
show a cusplike pattern of the specific heat, whose peak
 makes possible a better determination of 
the critical temperature. By increasing the lattice
dimension up to $N=15\times 15\times 15$,
the cusp becomes more pronounced, 
at variance with the case of the $2d$ model.
Fig. \ref{calspec.3d.fig} shows that this occurs
at the temperature $T_{c}\simeq2.17$.

\medskip
\subsubsection{Vorticity}
\medskip
The definition of the vorticity in the $3d$ case is not a simple extension
of the $2d$ case.
Vortices are always defined on a plane and if all the ``spins'' could freely
move
in the three-dimensional space, the concept of vortices would be meaningless.
For the $3d$ planar (anisotropic) model considered here, vortices can be
defined and studied on two-dimensional
subspaces of the lattice. The variables $q_{i,j,k}$ do not contain any
information about the position of the plane where the reference direction
to measure the angles $q_{i,j,k}$ is assigned.
Dynamics is completely independent of this choice, which has no effect on the
Hamiltonian. Moreover, as the Hamiltonian is symmetric with respect to the 
lattice axes, the three coordinate-planes are equivalent.  This 
equivalence implies that vortices can contemporarily exist on three 
orthogonal planes. Though the usual pictorial representation of a vortex
can hardly be maintained, its mathematical definition is the same as in the 
$2d$ lattice case. Hence three vorticity functions exist and their 
average values - at a given temperature - should not differ, what is actually
confirmed by numerical simulations.

\begin{figure}
\centerline{\includegraphics[width=0.80\linewidth]{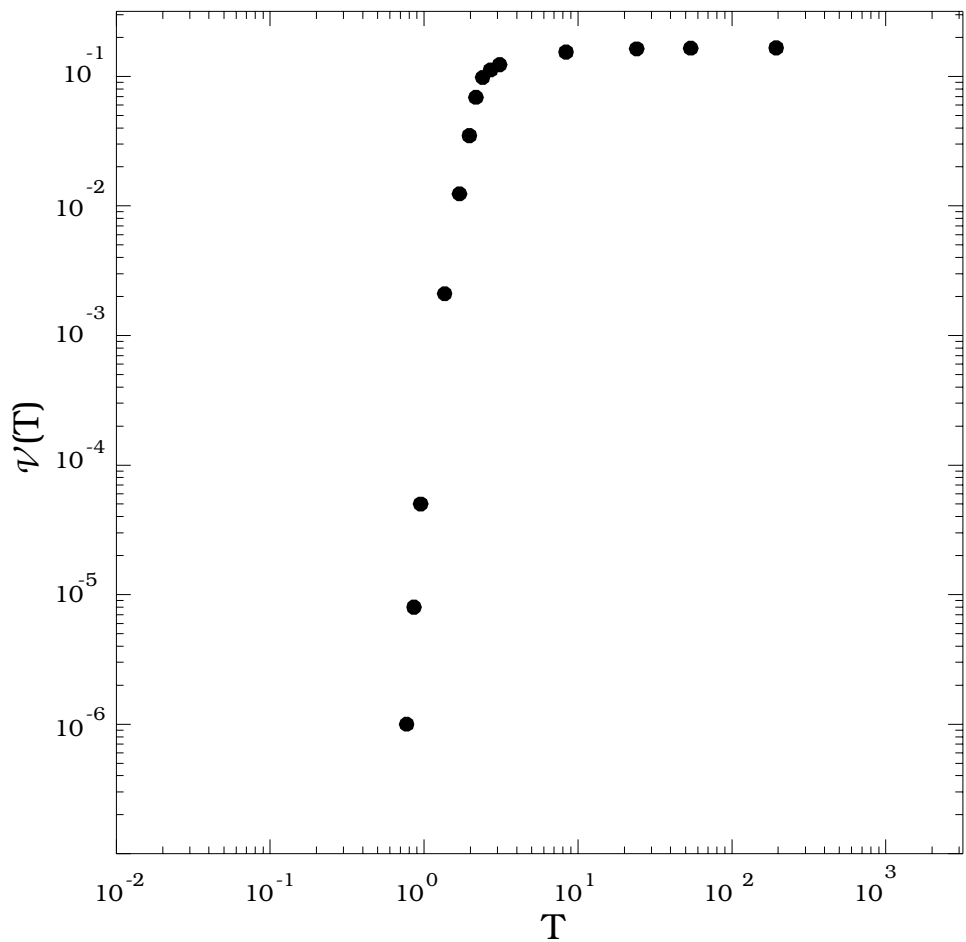}}
\caption{ Vorticity function at different temperatures along a dynamical
trajectory on a lattice of $N = 10 \times 10 \times 10$ sites. }
\label{vort.fig3d}
\end{figure}

The vorticity function vs. temperature is plotted in 
Fig. \ref{vort.fig3d}.
On a lattice of $10\times 10\times 10$ spins, the first vortex is  
observed at $T \simeq0.8$.
The growth of the average density of vortices is very
fast up to the critical temperature, above which the saturation is reached.

\medskip
\subsection{Lyapunov exponents and symmetry-breaking phase transition}
\medskip
A quantitative analysis of the dynamical chaoticity is provided by the 
temperature dependence  
of the largest Lyapunov exponent. 

\begin{figure}
\centerline{\includegraphics[width=0.80\linewidth]{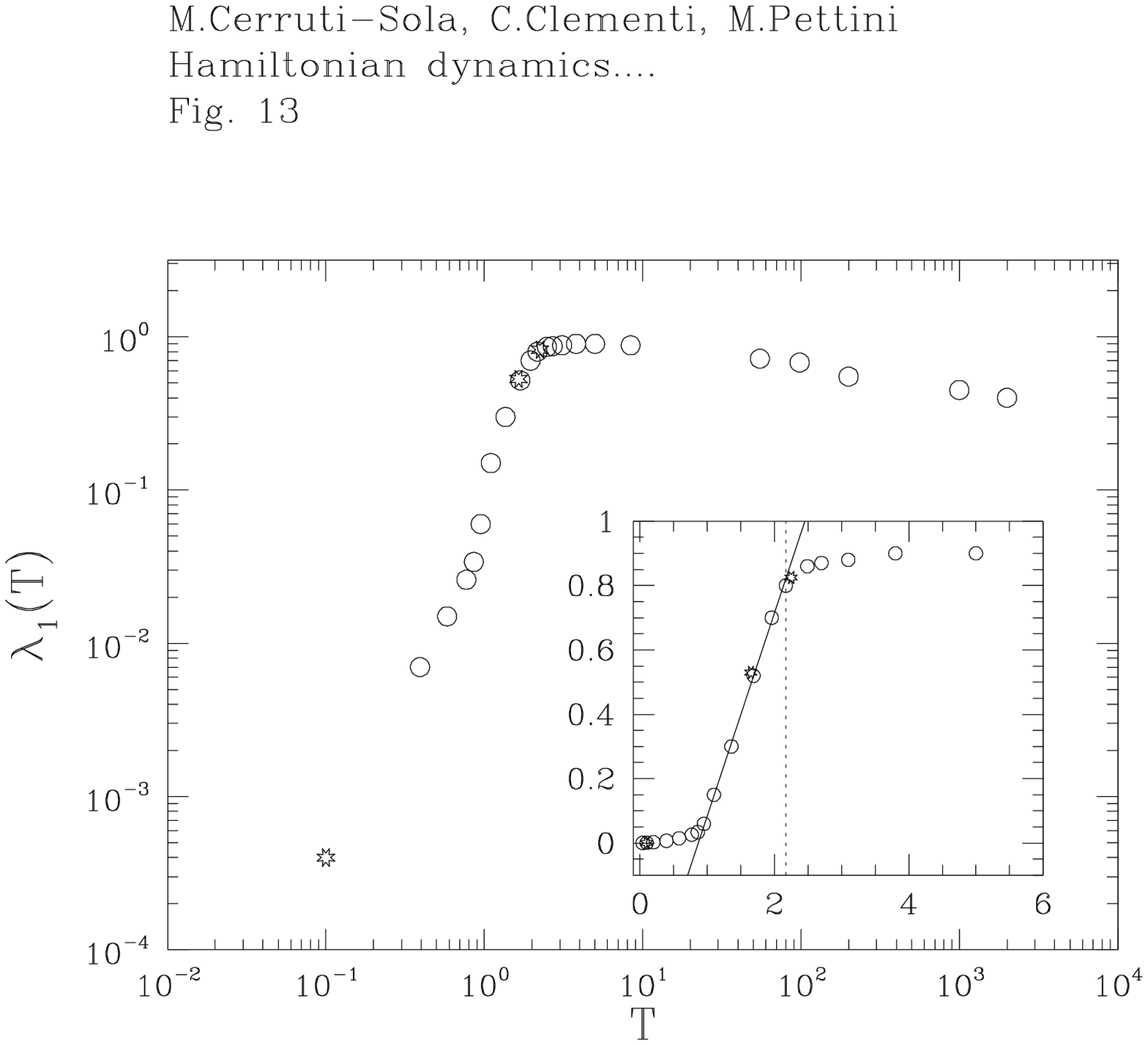}}
\caption{ The largest Lyapunov exponents computed at different temperatures
for the $3d$ model.
Numerical results are for lattices of $N= 10 \times 10 \times 10$ (open 
circles) and $N= 15 \times 15 \times 15$ (open stars). 
In the inset, symbols have the same meaning. The dashed line points out the
temperature $T_c \simeq 2.17$ of the phase transition. The solid line puts in
evidence the departure of $\lambda_1(T)$ from the linear growth. }
\label{lyap.3d.fig}
\end{figure}

Figure \ref{lyap.3d.fig} shows the results of this computation. 
At low temperatures, in the limit of quasi-harmonic oscillators, the scaling
law is again found to be
 $\lambda_{1}(T)\sim T^{2}$ and, at high
temperatures, the scaling law is again $\lambda_{1}(T)\sim T^{-1/6}$,  
as in the $2d$ case.
In the temperature range intermediate between $T \simeq0.8$ and 
$T_c \simeq2.17$, there is a linear growth of $\lambda_1(T)$.
At the critical temperature, the Lyapunov exponent exhibits an angular 
point. This makes a remarkable difference between this
system undergoing a second order phase transition and 
its $2d$ version, undergoing a Kosterlitz-
Thouless transition. In fact, the analysis of the $2d$ model has
shown a mild transition between the different regimes of $\lambda_{1}(T)$
(inset of Fig. \ref{vort.fig3d}),
whereas in the $3d$ model this transition 
is sharper (inset of Fig. \ref{lyap.3d.fig}).

We have also computed the temperature dependence of the largest Lyapunov 
exponent of Markovian random processes which replace the true dynamics on the 
energy surfaces $\Sigma_E$ (see Appendix).
The results are shown in Fig. \ref{randyn.3d.fig}.  
The dynamics is considered
strongly chaotic in the temperature range where the patterns  
$\lambda_1(T)$ are the same for both random and 
differentiable dynamics, i.e. when differentiable dynamics mimics,
to some extent,
a random process.
The dynamics is considered weakly chaotic when the value 
$\lambda_1$ resulting from
random dynamics is larger than the value $\lambda_1$
resulting from differentiable
dynamics.
 The transition from weak to strong chaos is quite abrupt. 
Figure \ref{randyn.3d.fig}  shows that 
the pattern of the largest Lyapunov exponent 
computed by means of the random dynamics 
reproduces that of the true Lyapunov exponent
at temperatures $T \geq T_c$. This means that the setting in
of strong thermodynamical disorder corresponds to the setting in of strong
dynamical chaos. 
 The ``window'' of strong chaoticity starts at
$T_c$ and ends at $T \sim 10$.
The existence of a second transition from strong to weak chaos is
due to the existence, for $T\rightarrow\infty$,
of the second integrable limit (of free 
rotators), whence chaos cannot remain strong at any $T>T_c$.

\medskip
\section{Geometry of dynamics and phase transitions}
\medskip
Let us briefly recall that the geometrization of the dynamics 
of $N$-degrees-of-freedom systems defined by a Lagrangian
${\cal L} = K - V$, in which the kinetic energy is quadratic in the velocities:
$K=\frac{1}{2}a_{ij} \dot{q}^i\dot{q}^j~$, stems from the fact that
the natural motions are the extrema
of the Hamiltonian action functional ${\cal S}_H = 
\int {\cal L} \, dt$, 
or of the Maupertuis' action
${\cal S}_M = 2 \int K\, dt$.
In fact, also the geodesics of Riemannian and pseudo-Riemannian 
manifolds are the extrema of a functional, the arc-length 
$\ell=\int ds$, with $ds^2={g_{ij}dq^i dq^j}$. 
Hence, a suitable choice of the metric tensor allows for the 
identification of the arc-length with either ${\cal S}_H$ or 
${\cal  S}_M$, and of the geodesics with the natural motions of the
dynamical system. Starting from ${\cal  S}_M$, the ``mechanical manifold''
is the accessible configuration space endowed with
the Jacobi metric \cite{Pettini} 
\beq
(g_J)_{ij} = [E - V(q)]\,a_{ij}~~,
\label{jacobi_metric}
\eeq 
where $V(q)$ is the potential energy and $E$ is the total energy.
A description of the extrema of Hamilton's 
action ${\cal S}_H$ as geodesics of a ``mechanical manifold'' 
can be obtained using Eisenhart's metric 
\cite{Eisenhart} on an enlarged configuration spacetime 
($\{q^0\equiv t,q^1,\ldots,q^N\}$ 
plus one real coordinate $q^{N+1}$), whose arc-length is
\begin{equation}
ds^2 = -2V(\{ q \}) (dq^0)^2 + a_{ij} dq^i dq^j + 2 dq^0 
dq^{N+1}~~.
\label{ds2E}
\end{equation}
The manifold has a Lorentzian structure and the dynamical 
trajectories are those geodesics satisfying the condition
$ds^2 = C dt^2$, where $C$ is a positive constant. 
In the geometrical framework, the (in)stability 
of the trajectories is the (in)stability 
of the geodesics, and it is completely determined by the 
curvature properties of the underlying manifold according to
the Jacobi equation \cite{Pettini,doCarmo}
\begin{equation}
\frac{\nabla^2 \xi^i}{ds^2} + R^i_{~jkm}\frac{dq^j}{ds} \xi^k 
\frac{dq^m}{ds} = 0~~,
\label{eqJ}
\end{equation}
whose solution $\xi$, usually called Jacobi or geodesic variation field,  
locally measures the distance between nearby geodesics; 
$\nabla/ds$ stands for the covariant derivative
along a geodesic and $R^i_{~jkm}$ are the components of 
the Riemann curvature tensor. 
Using the Eisenhart metric (\ref{ds2E}),
the relevant part of the Jacobi equation 
(\ref{eqJ}) is  \cite{CCP}
\begin{equation}
\frac{d^2 \xi^i}{dt^2} + R^i_{~0k0}\xi^k = 0~~,~~~~i=1,\dots ,N
\label{eqdintang}
\end{equation}
where the only non-vanishing components of the curvature tensor are
$R_{0i0j}=\partial^2 V/\partial q_i \partial q_j $. Equation 
(\ref{eqdintang}) is the tangent dynamics equation, which is commonly used to
measure Lyapunov exponents in standard Hamiltonian systems. 
Having recognized its geometric origin, 
it has been 
devised in Ref.\cite{CCP} a geometric reasoning
 to derive from  Eq.(\ref{eqdintang})
an {\it effective} scalar stability equation that, {\it independently} of the
knowledge of dynamical trajectories, provides an average measure of their
degree of instability. 
An intermediate step in this derivation yields
\beq
\frac{d^2 \xi^j}{dt^2} + k_R(t) \xi^j + \delta K^{(2)}{(t)} \xi^j = 0~~,
\label{sectional}
\eeq
where $k_R=K_R/N$ is the Ricci curvature along a geodesic defined as
$K_R = \frac{1}{v^2} R_{ij} {\dot{q}^i}{\dot{q}^j}$,
with $v^2 = {\dot{q}^i}{\dot{q}_i}$ and 
$R_{ij} = R^{k}_{~ikj}$, and $\delta K^{(2)}$ is the local
 deviation of sectional
curvature from its average value \cite{CCP}.
The sectional curvature is defined as $K^{(2)} = R_{~ijkl} \xi^i \dot{q}^j
\xi^k \dot{q}^l/ \parallel\xi\parallel^2 \parallel\dot{q}\parallel^2$.

Two simplifying assumptions are made:
$(i)$ the ambient manifold is {\em almost isotropic}, i.e. 
the components of the curvature tensor --- that for an isotropic manifold
(i.e. of constant curvature) 
are $R_{ijkm}=k_0(g_{ik} g_{jm} - g_{im} g_{jk})$, $k_0=const$ 
-- can be approximated by 
$R_{ijkm} \approx k(t)
(g_{ik} g_{jm} - g_{im} g_{jk})$
along a generic geodesic $\gamma(t)$; $(ii)$
in the large $N$ limit, the  ``effective curvature''
$k(t)$ can be modeled by a gaussian and $\delta$-correlated stochastic 
process.
Hence, one derives
 an effective
stability equation, independent of the dynamics and in the
form of a stochastic oscillator equation
\cite{CCP},
\begin{equation}
\frac{d^2\psi}{dt^2} + [k_0 + \sigma_k \eta(t)] \, \psi = 0~~,
\label{eqpsi}
\end{equation}
where $\psi^2 \propto |\xi|^2$. 
 The mean $k_0$ and variance $\sigma_k$ of $k(t)$ 
are given by  
$k_0 =  {\langle K_R \rangle}/{N}$
and $\sigma^2_k  =  {\langle (K_R - \langle K_R \rangle)^2 \rangle}/{N}$,
respectively, and the averages $\langle\cdot\rangle$ are geometric averages,
i.e. integrals computed on the mechanical manifold. These averages are 
directly related with microcanonical averages, as it will be seen at the end
of Section V.
$\eta(t)$ is a gaussian $\delta$-correlated random process of
zero mean and unit variance.

The main source of instability of the solutions of Eq.(\ref{eqpsi}),
and therefore the main source of Hamiltonian chaos, is 
parametric resonance, which is
 activated by the variations of the Ricci curvature
along the geodesics and which takes place also on positively curved manifolds
\cite{CerrutiPettini}. The dynamical instability can be enhanced 
if the geodesics encounter regions 
of negative sectional curvatures, such that $k_R + \delta K^{(2)}
< 0$, as it is evident from Eq. (\ref{sectional}).

In the case of Eisenhart metric, 
it is 
$K_R\equiv \Delta V = \sum_{i=1}^N ({\partial^2 V}/{\partial q_i^2})$
and $K^{(2)} = R_{~0i0j} \xi^i \xi^j/ \parallel \xi \parallel^2 \equiv
(\partial^2V/\partial q^i\partial q^j)\xi^i\xi^j/\Vert\xi\Vert^2$.
The exponential growth rate $\lambda$ of the 
quantity $\psi^2+\dot\psi^2$ of the solutions of Eq. (\ref{eqpsi}),
is therefore an estimate of the largest Lyapunov exponent that can be 
analytically computed. The final result reads \cite{CCP} 
\begin{equation}
\lambda = \frac{\Lambda}{2} - \frac{2 k_0}{3 \Lambda}\,,~~
\Lambda = \left(2\sigma_k^2 \tau +
\sqrt{\frac{64 k_0^3}{27} + 4\sigma_k^4 \tau^2}~\right)^\frac{1}{3}~,
\label{lambda}
\end{equation}
where
$\tau = \pi\sqrt{k_0}/(2\sqrt{k_0(k_0 + \sigma_k)}
+\pi\sigma_k )$;
in the limit $\sigma_k/k_0 \ll 1$ one finds $\lambda \propto \sigma_k^2~$.

\subsection{Signatures of phase transitions from geometrization of 
dynamics}

In the geometric picture, chaos is mainly originated by the parametric 
instability activated by the fluctuating curvature felt by geodesics,
i.e. the fluctuations of the (effective)
curvature are the source of the instability of the dynamics.
On the other hand, as it is witnessed by the derivation of Eq. (\ref{eqpsi})
and by the equation itself, a statistical-mechanical-like treatment of the
average degree of chaoticity is made possible by the geometrization of the
dynamics. The relevant curvature properties of the mechanical manifolds
are computed, at the formal level, as statistical averages, like other
thermodynamic observables. Thus, we can expect that some precise relationship
may exist between geometric, dynamic and thermodynamic quantities.
Moreover, this implies that phase transitions should correspond to peculiar
effects in the geometric observables.

In the particular case of the $2d$ XY model, 
the microcanonical average kinetic energy $\langle K \rangle$
and the average Ricci curvature 
$\langle K_R \rangle$ computed 
with the Eisenhart metric are linked by the equation 
\beq
K_{R}=
\left \langle \sum _{i,j=1}^{N}\frac{\partial^{2} V}{\partial^2 q_{i,j}}
\right \rangle = 2J
\sum_{i,j=1}^{N}
\left\langle\cos(q_{i+1,j}-q_{i,j})
+\cos(q_{i,j+1}-q_{i,j}) \right\rangle =2(J-\langle V \rangle )~~,
\label{kinRicci}
\eeq
so that
\beq
H=N\epsilon=\langle K \rangle + \langle V \rangle \mapsto\frac{ \langle K 
\rangle }{N}=\epsilon-2J+\frac{1}{2}\frac{\langle K_{R} \rangle}{N}~~.
\label{cRicci.Uinterna}
\eeq
Being the temperature defined as $T=2\langle K \rangle /N$
(with $k_{B}=1$) and being $d=1$ (because each spin has only one
rotational degree of freedom), from Eq.(\ref{specheat}) it follows that

\beq
c_{V}=\frac{1}{2}\left(1-\frac{1}{2}\frac{\sigma^{2}_k/N}{T^2}\right)^{-1}.
\label{fluttKr.calspec}
\eeq

In the special case of these XY systems,
it is possible to link the specific heat and the Ricci curvature
by inserting Eq.(\ref{cRicci.Uinterna}) into the usual expression for the 
specific heat at constant volume. Thus, one obtains the equation
\beq
c_{V}=-\frac{1}{2N}\frac{\partial \langle K_{R} \rangle(T)}{\partial T}~.
\label{calspec,Ricci}
\eeq
The appearance of a peak in the specific heat function at the critical 
temperature
has to correspond to a suitable temperature dependence of the Ricci curvature.

In the $3d$ model, the
potential energy and the Ricci curvature 
are proportional, according to:
$\frac{1}{N} \langle V \rangle = 3 - \frac{1}{2 N} \langle K_{R} \rangle~$.

Another interesting point is the relation between a geometric observable
and the vorticity function in both models.
As already seen in previous sections, the
vorticity function is a useful signature of the dynamical chaoticity 
of the system. From the geometrical point of view,
the enhancement of the instability of the dynamics
with respect to the parametric instability due to curvature fluctuations,
is linked
to the probability of obtaining negative sectional curvatures 
along the geodesics (as discussed for $1d$ XY model in Ref.\cite{CCP}).
In fact, when vortices are present in the system, 
there will surely be two neighbouring spins with an orientation difference 
greater than $\pi/2$, such that, if $i,j$ and $i+1,j$ are their coordinates
on the lattice, it follows that
\beq
q_{i+1,j}-q_{i,j} > \frac{\pi}{2} \rightarrow \cos(q_{i+1,j}-q_{i,j})<0~.
\label{coord}
\eeq
The sectional curvature relative to the plane defined 
by the velocity ${\bf v}$ along
a geodesic and a generic vector  ${\bf \xi}\perp {\bf v}$ is 
\beq
K^{(2)}= \sum_{i,j,k,l=1}^{N} \frac{\partial^2V}
{\partial q_{i,j}\partial q_{k,l}}
\frac{\xi^{i,j}\xi^{k,l}}{\|{\bf \xi}\|^{2}}~~.
\label{kappa2}
\eeq
For the $2d$ XY model, it is 
\beq
K^{(2)}= \frac{J}{\|{\bf \xi}\|^{2}}\sum_{i,j=1}^{N}\{\cos(q_{i+1,j}-q_{i,j})
[\xi^{i+1,j}-\xi^{i,j}]^{2} +\cos(q_{i,j+1}-q_{i,j})[\xi^{i,j+1}-\xi^{i,j}]^{2}
\}~.
\label{curvsezXY}
\eeq
Thus, a large probability of
having a negative value of the cosine of the difference among the directions
of two close spins corresponds to a larger probability of obtaining
negative values of the sectional curvatures along the geodesics; here for $\xi$
the geodesic separation vector of Eq.(\ref{eqdintang}) is chosen.

In the $3d$ model, the sectional curvature relative to
the plane defined by the velocity ${\bf v}$ and 
a generic vector ${\bf \xi}\perp {\bf v}$ is 
\begin{eqnarray}
K^{(2)}&=& \frac{J}{\|\xi\|^{2}}\sum_{i,j,k=1}^{N} \{
\cos(q_{i+1,j,k}-q_{i,j,k})[\xi^{i+1,j,k}-\xi^{i,j,k}]^{2}+
\nonumber\\
&+&\cos(q_{i,j+1,k}-q_{i,j,k})[\xi^{i,j+1,k}-\xi^{i,j,k}]^{2}+
\cos(q_{i,j,k+1}-q_{i,j,k})[\xi^{i,j,k+1}-\xi^{i,j,k}]^{2}\}
\label{curvsez3d}
\end{eqnarray}
and again the probability of finding negative values of $K^{(2)}$
along a trajectory is limited to the probability of finding vortices.

The mean values of the geometric quantities entering Eq.(\ref{eqpsi})
can be numerically computed by means of Monte Carlo simulations
or by means of time averages along the
dynamical trajectories. In fact, due to the lack of an explicit expression
for the canonical partition function of the system,
these averages are not analytically computable.
For sufficiently high
temperatures, the potential energy becomes 
negligible with respect to the kinetic energy, and each spin is free to move
independently from the others. Thus, in the limit of high temperatures,
one can estimate the configurational partition function
$Z_C = \int_{-\pi}^{\pi} \prod_{\bf i} dq_{\bf i} e^{-\beta V(q)}$ 
by means of the expression  
\begin{eqnarray}
Z_C& =& e^{-2 \beta JN}
\int_{-\pi}^{\pi} \prod_{i,j=1}^{N}
dq_{i,j}\exp\{\beta J \sum_{i,j=1}^{N}[
\cos(q_{i+1,j}-q_{i,j})+\cos(q_{i,j+1}-q_{i,j})]\}\nonumber\\
&\sim& e^{-2 \beta JN}\int_{-\pi}^{\pi} \prod_{i,j=1}^{N}du_{i,j}dv_{i,j}
\exp\{\beta J \sum_{i,j=1}^{N}[
\cos(u_{i,j})+\cos(v_{i,j})]\}
\label{partition}
\end{eqnarray}
after the introduction of $u_{i,j}= q_{i+1,j}-q_{i,j}$ and 
$v_{i,j}= q_{i,j+1}-q_{i,j}$ as independent variables.
In this way, some analytical estimates of the average Ricci
curvature $k_{0}(T)$ and of its r.m.s. fluctuations 
$\sigma^2_k(T)$ have been obtained for the $2d$ model
 (Fig. \ref {fig.Ricci2d}).
For temperatures above the temperature of the Kosterlitz-Thouless
transition, these estimates
are in agreement
with the numerical computations on a $N=10\times 10$ lattice.
It is confirmed that Hamiltonian dynamical simulations,
already on rather small lattices, 
are useful to predict, with a good approximation, the thermodynamic limit
behavior of relevant observables.
Moreover, the good quality of the high temperature estimate gives
a further information: at the transition temperature,
the correlations among the different degrees of freedom are destroyed,
confirming the
strong chaoticity of the dynamics.

The same high temperature estimates of $k_0(T)$ and $\sigma^2_{k}(T)$
have been performed for the $3d$ system.
In Fig. \ref{keflutt.3dfig},
the numerical determination of $\sigma^2_{k}(T)$
shows the appearance of a very pronounced peak at the phase transition point
which is not predicted by the analytic estimate, 
whereas the average Ricci curvature $k_0(T)$ is in agreement
with the analytic values of the high temperature estimate, computed by spin 
decoupling, above the critical temperature, as in the $2d$ model.

\medskip
\subsection{Geometric observables and Lyapunov exponents}
\medskip
We have seen that the largest
Lyapunov exponent is sensitive to the phase transition and at the same time
we know that
it is also related to the average curvature properties of the ``mechanical
manifolds''.
Thus, the geometric observables $k_0(T)$ and $\sigma^2_k(T)$
above considered can be used to 
estimate the Lyapunov exponents, as well as to detect the phase transition.

In principle, by means of Eq.(\ref{lambda}), one can evaluate
the largest Lyapunov exponent without any need of dynamics,
but simply using 
global geometric quantities of the manifold 
associated to the physical system.
For $2d$ and $3d$ XY models, fully analytic computations are possible only
in the limiting cases of high and low temperatures. Microcanonical averages
of $k_0$ and $\sigma^2_k$ at arbitrary $T$ have been numerically computed
through time averages. We can call this hybrid method semi-analytic.

In Fig. \ref{prev.Lyap.2d}, the results of the semi-analytic prediction of the 
Lyapunov exponents for the $2d$ model are plotted vs. temperature and 
compared with the numerical outcomes of the tangent dynamics.
As one can see, the prediction formulated on the basis of Eq.(\ref{lambda}) 
underestimates the numerical values given by the tangent 
dynamics.
The semi-analytic prediction can be improved
by observing that the replacement of the sectional curvature fluctuation
$\delta K^{(2)}$ in Eq.(\ref{sectional}) with a fraction 
of the Ricci curvature [which underlies the derivation of Eq.(\ref{eqpsi})] 
underestimates the frequency of occurrence of negative sectional curvatures,
which was already the case of the $1d$ XY model \cite{CCP}.
The correction procedure can be implemented
by evaluating the probability $P(T)$ of obtaining a negative value of the 
sectional curvature along a generic trajectory and then by operating the 
substitution
\beq
K_{R}(T)\rightarrow \frac{K_{R}(T)}{1+P(T) \alpha}~.
\label{kappafrac}
\eeq
The parameter $\alpha$ is a free parameter to be empirically estimated.
Its value ranges from $100$ to $200$, without appreciable differences in the
final result.
It resumes 
the non trivial information about the more pronounced tendency
of the trajectories towards negative
sectional curvatures with respect to the predictions of the geometric model
describing the chaoticity of the dynamics.

The probability $P(T)$ is estimated through the occurrence along 
a trajectory of negative values 
of the sum of the coefficients that appear in the definition
of $K^{(2)}$ [Eqs.(\ref{curvsezXY}) and (\ref{curvsez3d})]
\beq
P(T)\sim \frac{\int_{-\pi}^{\pi}
\Theta (-\cos(q_{k+1,l}-q_{k,l})-\cos(q_{k,l+1}-q_{k,l}))
\exp[-\beta V({\bf q})] \prod_{k,l=1}^{N}dq_{k,l}}
{\int_{-\pi}^{\pi}\exp[-\beta V({\bf
q})] \prod_{k,l=1}^{N}dq_{k,l}}~~,
\label{2d.pro.cos.neg}
\eeq
averaged over all the sites
$\forall k,l\in (1,\ldots,N)$; $\Theta$ is the step function.

Alternatively, owing to the already remarked relation between vorticity
and sectional curvature $K^{(2)}$,
$P(T)$ can be replaced by the average density of
vortices
\beq
K_{R}(T)\rightarrow \frac{K_{R}(T)}{1+\overline{\alpha} {\cal V}(T)}~,
\label{correction}
\eeq
where $\overline{\alpha}$ a free parameter. 
Actually,
in the $2d$ model, the two corrections, one given by
Eq.(\ref{kappafrac}) with $P(T)$ of Eq. (\ref{2d.pro.cos.neg}), the other 
given by Eq.(\ref{correction}) with the vorticity function in place of $P(T)$,
convey the same information.
The semi-analytic predictions of $\lambda_1(T)$ with correction
are reported in  Fig. \ref{prev.Lyap.2d}.

In the limits of high and low temperatures, 
$\lambda_1(T)$ can be given the analytic forms 
$\lambda_{1}(T)\sim T^{-1/6}$ at high 
temperature, and 
$\lambda_{1}(T)\sim T^{2}$ at low temperature.
In the former case, the high temperature approximation (\ref{partition})
is used, and in the latter case the quasi-harmonic oscillators approximation
is done. The deviation of $\lambda_1(T)$ from the quasi-harmonic scaling,
starting at $T \simeq0.6$ and already observed to correspond to the 
appearance of vortices, finds here a simple explanation through the geometry
of dynamics: vortices are associated with negative sectional curvatures,
enhancing chaos.

By increasing the spatial dimension of the system, it becomes more and more
difficult to accurately estimate the probability of obtaining negative
sectional curvatures. 
The assumption that the occurrence of negative values
of the cosine of the difference between the directions of two nearby spins
is nearly equal to $P(T)$, is less effective in the $3d$ 
model than in the $2d$ one. 
Again, the vorticity function can be assumed as an estimate of $P(T)$
[Eq. (\ref{correction})].
The quality of the results has a weak dependence upon  
the parameter $\alpha$.
The correction remains good, with $\alpha$ belonging to a broad interval
of values ($100 \div 200$). 
In the limits of high and low temperatures, the model predicts correctly
the same scaling laws of the $2d$ system.

In Fig. \ref{prev.Lyap.3d} the semi-analytic
 predictions for the Lyapunov exponents, 
with and without correction, are 
plotted vs. temperature together with the numerical results
of the tangent dynamics.
It is noticeable that the prediction of Eq. (\ref{lambda}) is able to give
the correct asymptotic behavior of the Lyapunov exponents
also at low temperatures, the most difficult part 
to obtain by means of dynamical simulations. 

\medskip
\subsection{A topological hypothesis}
\medskip

We have seen in Fig. \ref{keflutt.3dfig} that a sharp peak of the 
Ricci-curvature fluctuations
$\sigma_\kappa^2(T)$ is found for the $3d$ model in correspondence of the 
second order phase transition, whereas, for the $2d$ model, 
$\sigma_\kappa^2(T)$ appears regular and in agreement with the theoretically 
predicted smooth pattern.
On the basis of heuristic arguments, in Refs.\cite{CCCP,CCCPPG} we suggested
that the peak of $\sigma_\kappa^2$ observed for the $3d$ XY model, as well as
for $2d$ and $3d$ scalar and vector lattice $\varphi^4$ models, might originate
in some change of the {\it topology} of the mechanical manifolds. In fact,
in abstract mathematical models, consisting of families of surfaces undergoing
a topology change -- i.e. a loss of diffeomorphicity among them -- 
at some critical value of
a parameter labelling the members of the family, we have actually observed the
appearance of cusps of $\sigma_K^2$ at the 
transition point between two subfamilies of surfaces of different topology,
$K$ being the Gauss curvature.

Actually, for the mean-field XY model, where both $\sigma_\kappa^2(T)$ and 
$\lambda_1(T)$ have theoretically been shown to loose analyticity at the phase 
transition point, a direct evidence of a ``special'' change of the topology
of equipotential hypersurfaces of configuration space has been given
\cite{cegdcp}. Other indirect and direct evidences of the actual involvement
of topology in the deep origin of phase transitions have been recently
given 
\cite{fps1,fps2} for the lattice $\varphi^4$ model.

In the following Section we consider the extension 
of this
topological point of view about phase transitions
from equipotential hypersurfaces of configuration space
to constant energy hypersurfaces of phase space.

\medskip
\section{Phase space geometry and thermodynamics.}
\medskip

In the preceding Section we have used some elements of intrinsic differential
geometry of submanifolds of configuration space to describe the average
degree of dynamical instability (measured by the largest Lyapunov exponent).
In the present Section we are interested in the relationship
between the extrinsic geometry of the constant energy
hypersurfaces $\Sigma_E$ and thermodynamics.

Hereafter, phase space is considered as an even-dimensional subset $\Gamma$
of ${\Bbb R}^{2N}$
and the hypersurfaces $\Sigma_E=\{(p_1,\dots ,p_N,q_1,\dots ,q_N)\in{\Bbb R}
\vert H(p_1,\dots ,p_N,q_1,\dots ,q_N)=E\}$ are manifolds that can be equipped
with the standard Riemannian metric induced from ${\Bbb R}^{2N}$. If, for 
example, a surface is parametrically defined through the equations
$x^i=x^i(z^1,\dots ,z^k)$, $i=1,\dots ,2N$, then the metric $g_{ij}$
{\it induced} on the surface is given by $g_{ij}(z^1,\dots ,z^k)=
\sum_{n=1}^{2N}
\frac{\partial x^n}{\partial z^i} \frac{\partial x^n}{\partial z^j}$. 
The geodesic flow associated with the metric induced on $\Sigma_E$ from
${\Bbb R}^{2N}$ has nothing to do with the 
Hamiltonian flow that belongs to $\Sigma_E$. Nevertheless, it exists an
intrinsic 
Riemannian metric $g_S$ of phase space $\Gamma$ such that the geodesic 
flow of $g_S$, restricted to $\Sigma_E$, coincides with the Hamiltonian flow
($g_S$ is the so called Sasaki lift to the tangent bundle
of configuration space of the Jacobi
metric $g_J$ that we mentioned in a preceding Section).

The link between extrinsic geometry of the $\Sigma_E$ and thermodynamics
is estabilished through the  microcanonical definition of entropy
\beq
S = k_B \log \int_{\Sigma_E} \frac{d\sigma}{\|{\nabla H}\|}  ~,
\label{entropy}
\eeq
where $d\sigma = \sqrt{det(g)} dx_1...dx_{2N-1}$ is the invariant volume
element of $\Sigma_E \subset{\Bbb R}^{2N}$, ${g}$ is the metric induced from
${\Bbb R}^{2N}$ and $x_1...x_{2N-1}$ are the coordinates on $\Sigma_E$.

Let us briefly recall some necessary definitions and concepts that are needed
in the study of hypersurfaces of euclidean spaces.

A standard way to investigate the geometry of an hypersurface $\Sigma^m$ is 
to study the way in which it curves around in ${\Bbb R}^{m+1}$: this is 
measured by the way the normal direction  changes as we move from point to
point on the surface. The rate of change
of the normal direction ${\bf N}$ at a point $x\in\Sigma$ is described by
the {\it shape operator} $L_x({\bf v}) = - \nabla_{\bf v}{\bf N}
 = - (\nabla N_1\cdot{\bf v},\dots ,\nabla N_{m+1}
\cdot{\bf v})$, where
${\bf v}$ is a tangent vector at $x$ and $\nabla_{\bf v}$ is the directional
derivative of the unit normal ${\bf N}$.
As $L_x$ is an operator of the tangent space at $x$ into
itself, there are $m$ independent eigenvalues \cite{thorpe} $\kappa_1(x),
\dots,\kappa_m(x)$, which are called the principal curvatures of $\Sigma$ at
$x$. Their product is the {\it Gauss-Kronecker curvature}: 
$K_G(x)=\prod_{i=1}^m\kappa_i(x)={\rm det}(L_x)$, and
their sum is the so-called {\it mean curvature}:
$M_1(x)=\frac{1}{m}\sum_{i=1}^m\kappa_i(x)$.
The quadratic form $L_x({\bf v})\cdot{\bf v}$, associated with the shape 
operator at a point $x$, is called the second fundamental form of $\Sigma$ at 
$x$.

It can be 
 shown \cite{doCarmo} that the mean curvature of the energy hypersurfaces
 is given by
\begin{equation}
M_1(x)= -\frac{1}{2N-1}\nabla\cdot\left(\frac{\nabla H(x)}{\Vert\nabla H(x)
\Vert}\right)~~,
\label{emme1}
\end{equation} 
where $\nabla H(x)/\Vert\nabla H(x)\Vert$ is the unit normal to
$\Sigma_E$ at a given point $x=(p_1,\dots ,p_N,q_1,\dots ,q_N)$, and 
$\nabla =(\partial/\partial p_1,\dots,\partial/\partial q_N)$, 
whence the explicit expression
\begin{eqnarray}
(2N-1)\, M_1 = &-&\frac{1}{\Vert\nabla H\Vert}\left[ N + \sum_{\bf i}
\left(\frac{\partial^2V}{\partial q_{\bf i}^2}\right)\right]\nonumber\\
& +& \frac{1}{\Vert\nabla H\Vert^3}
\left[\sum_{\bf i}p_{\bf i}^2 + \sum_{\bf i,j}\left(
\frac{\partial^2V}{\partial q_{\bf i}\partial q_{\bf j}}\right)\left(
\frac{\partial V}{\partial q_{\bf i}}\right)\left(\frac{\partial V}
{\partial q_{\bf j}}\right)\right]~,
\label{M1}
\end{eqnarray}
where ${\bf i,j}$ are multi-indices according to the number of spatial 
dimensions.
 
The link between geometry and physics stems from the microcanonical definition
of the temperature
\begin{equation}
\frac{1}{T}=\frac{\partial S}{\partial E}=\frac{1}{\Omega_{\nu}(E)}\frac
{d\Omega_{\nu}(E)}{dE}~,
\label{temperature}
\end{equation}
where we used Eq.(\ref{entropy}) with $k_B=1$, $\nu =2N-1$, and 
$\Omega_{\nu}(E)=\int_{\Sigma_E}d\sigma /\Vert\nabla H\Vert$. 
>From the formula \cite{laurence} 
\begin{equation}
{d^k \over dE^k} \left( \int_{\Sigma_E} \alpha~d\sigma \right) (E')
        = \int_{\Sigma_{E'}} A^k(\alpha)\, d\sigma~~,
\end{equation}
where $\alpha$ is an integrable function and $A$ is the operator 
$        A(\alpha)= {\nabla \over  \Vert \nabla H\Vert}
         \cdot \left( \alpha \cdot
        {\nabla H \over \Vert \nabla H\Vert} \right),
$
it is possible to work out the result
\begin{equation}
\frac{1}{T}=
\frac{1}{\Omega_{\nu}(E)}\frac{d\Omega_{\nu}(E)}{dE} = \frac{1}{\Omega_{\nu}}
\int_{\Sigma_E} \frac{d\sigma}{\Vert\nabla
H\Vert} \left[ 2 \frac{M_1^\star}{\Vert\nabla H\Vert} - \frac{\triangle H}
{\Vert\nabla H\Vert^2} \right]\simeq \frac{1}{\Omega_{\nu}} 
\int_{\Sigma_E} \frac{d\sigma}
{\Vert\nabla H\Vert}\, \frac{M_1^\star}{\Vert\nabla H\Vert} ~~ ,
\label{ballerotte}
\end{equation}
where $M_1^\star = \nabla (\nabla H/ \Vert\nabla H\Vert)$ is directly
proportional to the mean curvature (\ref{emme1}). In the last term of
Eq.(\ref{ballerotte}) we have neglected a contribution which vanishes
as ${\cal O}(1/N)$. Eq. (\ref{ballerotte}) provides the
fundamental link between extrinsic geometry and thermodynamics \cite{nota1}. 
In fact, the microcanonical average of ${M_1^\star}/{\Vert\nabla H\Vert}$, 
which is a quantity 
tightly related with the mean curvature of $\Sigma_E$, gives the inverse
of the temperature, whence other important thermodynamic observables can be 
derived. For example, the constant volume specific heat
\begin{equation}
\frac{1}{C_V}=\frac{\partial T(E)}{\partial E}~,
\label{cv}
\end{equation}
using Eq.(\ref{temperature}), yields
\begin{equation}
C_V=-\left(\frac{\partial S}{\partial E}\right)^2\,\left(\frac{\partial^2S}
{\partial E^2}\right)^{-1}~~,
\label{cv1}
\end{equation}
becoming at large $N$ 
\begin{equation}
C_V=-\left<\frac{{M_1^\star}}{{\Vert\nabla H\Vert}}\right>_{mc}^2 \left[
 \frac{1}{\Omega_{\nu}} \frac{d}{dE}
\int_{\Sigma_E} \frac{d\sigma}
{\Vert\nabla H\Vert}\,\left( \frac{M_1^\star}{\Vert\nabla H\Vert}+R(E)\right)
- \left<\frac{{M_1^\star}}{{\Vert\nabla H\Vert}}\right>_{mc}^2\right]^{-1}~,
\label{cv2}
\end{equation}  
where the subscript $mc$ stands for microcanonical average, and
$R(E)$ stands for the quantities of order ${\cal O}(1/N)$  
neglected in the last term of Eq.(\ref{temperature})
(a-priori, its derivative can be non negligible and has to be taken into 
account).
Eq. (\ref{cv2}) highlights a more elaborated link between geometry and
thermodynamics: the specific heat depends upon the microcanonical average of
${M_1^\star}/{\Vert\nabla H\Vert}$ and upon the energy variation rate of the 
surface integral of this quantity. 

Remarkably, the relationship between curvature
properties of the constant energy surfaces $\Sigma_E$ and thermodynamic
observables given by Eqs.(\ref{temperature}) and (\ref{cv2}) can be extended
to embrace also a deeper and very interesting relationship between 
thermodynamics and {\it topology} of the constant
energy surfaces. Such a relationship can be discovered through a 
reasoning which, though approximate, is highly non-trivial, for it makes use
of a deep theorem due to Chern and Lashof \cite{ChernLashof}. 
As $\Vert\nabla H\Vert =\{\sum_i p_i^2 + [\nabla_iV(q)]^2\}^{1/2}$ is a
positive quantity increasing with the energy, we can write
\begin{equation}
\frac{1}{T}=
\frac{1}{\Omega_{\nu}}\frac{d\Omega_{\nu}}{dE} \simeq\frac{1}{\Omega_{\nu}}
\int_{\Sigma_E} \frac{d\sigma}
{\Vert\nabla H\Vert}\, \frac{M_1^\star}{\Vert\nabla H\Vert}
= D(E)\frac{1}{\Omega_{\nu}} \int_{\Sigma_E} d\sigma \, M_1~,
\label{deformaz}
\end{equation} 
where we have introduced the factor function $D(E)$ in order to extract the
total mean curvature $\int_{\Sigma_E} d\sigma \, M_1$; $D(E)$ has been 
numerically found to be smooth and very close to $\langle {1}/{\Vert
\nabla H\Vert^2} \rangle_{mc}$ (see Section \ref{5a} and Fig. 
\ref{D(E)}). Then,  
recalling the expression of a multinomial expansion  
\begin{equation}
        (x_1 + \cdots + x_{\nu})^{\nu} = \sum_{_{\{n_i\},\sum n_k =\nu}} 
\frac{\nu !}{n_1!\cdots n_{\nu}!} \cdot x_1^{n_1} \cdots x_{\nu}^{n_{\nu}}~~ ,
\label{somma1}
\end{equation}
and identifying the $x_i$ with the principal curvatures $k_i$, one obtains
\begin{equation}
 M_1^{\nu} = {\nu}! \prod_{i=1}^{\nu} k_i + R = {\nu}! K + R~~~,
\end{equation}
where $K=\prod_i k_i$ is the Gauss-Kronecker curvature,
and $R$ is the sum (\ref{somma1}) without the term with the
largest coefficient ($n_k=1, \ \forall k$).
Using $\nu ! \simeq \nu^{\nu} e^{-\nu} \sqrt{2\pi\nu}$, 
\begin{equation}
  M_1^{\nu} \simeq \nu^{\nu} e^{-\nu} \sqrt{4\pi N}  K + R
\label{pallino}
\end{equation}
is obtained. 
The above mentioned theorem of Chern and Lashof 
states that
\begin{equation}
\int_{\Sigma_E}
\vert K\vert\,d\sigma\geq Vol[{\Bbb S}_1^{\nu}]\sum_{i=0}^{\nu} 
b_i(\Sigma_E )~,
\label{chern1}
\end{equation}
i.e. the total absolute Gauss-Kronecker curvature of a hypersurface is
related with the sum of all its Betti numbers $b_i(\Sigma_E)$. The Betti 
numbers are {\it diffeomorphism invariants} of fundamental topological 
meaning \cite{Betti}, therefore their sum is also a topologic invariant.
${\Bbb S}_1^\nu$ is a hypersphere of unit radius. Combining Eqs.
(\ref{pallino}) and (\ref{chern1}) and integrating on $\Sigma_E$, we obtain
\begin{equation}
  \int_{\Sigma_E}\vert M_1^{\nu}\vert\, d \sigma \simeq   
\nu^{\nu} e^{-\nu} \sqrt{2\pi\nu} 
\int_{\Sigma_E}\vert K\vert d \sigma + \int_{\Sigma_E}\vert R\vert d \sigma
\ge {\cal A}\,\sum_{i=0}^{\nu} b_i(\Sigma_E ) + {\cal R}(E)~~,
\label{tria}
\end{equation}
with the shorthands        
${\cal A}=\nu^{\nu} e^{-\nu} Vol(S_1^{\nu})$ and  ${\cal R}=\int_{\Sigma_E}
\vert R\vert d \sigma$. 
 
Now, with the aid of the inequality $\int\Vert f\Vert^{1/n}d\mu\geq\Vert\int f
d\mu\Vert^{1/n}$, we can write
\begin{equation}
\int_{\Sigma_E}\vert M_1\vert\, d \sigma = \int_{\Sigma_E}\vert M_1^{\nu}\vert
^{1/\nu}\, d \sigma \geq  
\left\vert\int_{\Sigma_E}M_1^{\nu}\,d\sigma\right\vert^{1/\nu}~~.
\label{chern2}
\end{equation}
If $M_1\geq 0$ everywhere on $\Sigma_E$, then 
$\left\vert\int_{\Sigma_E}M_1^{\nu}\,d\sigma\right\vert^{1/\nu} =
\left(\int_{\Sigma_E}\vert M_1^{\nu}\vert\,d\sigma\right)^{1/\nu}$, whence, in 
the hypothesis that $M_1\geq 0$ is largely prevailing \cite{bassi-indici}, 
$\left\vert\int_{\Sigma_E}M_1^{\nu}\,d\sigma\right\vert^{1/\nu} \sim
\left(\int_{\Sigma_E}\vert M_1^{\nu}\vert\,d\sigma\right)^{1/\nu}$. Under the 
same assumption, $\int_{\Sigma_E}M_1d\sigma\sim\int_{\Sigma_E}\vert M_1\vert
d\sigma$ and therefore 
\begin{equation}
\int_{\Sigma_E} M_1\, d \sigma \sim \int_{\Sigma_E}\vert M_1^{\nu}\vert
^{1/\nu}\, d \sigma \geq  
\left\vert\int_{\Sigma_E}M_1^{\nu}\,d\sigma\right\vert^{1/\nu} \sim  
\left(\int_{\Sigma_E}\vert M_1^{\nu}\vert\,d\sigma\right)^{1/\nu} 
\geq \left[ {\cal A}\sum_{i=0}^{\nu} b_i(\Sigma_E ) + {\cal R}(E)
\right]^{1/\nu}~~.
\label{chern3}
\end{equation}
Finally,
\begin{eqnarray}
\frac{1}{T(E)} = 
\frac{1}{\Omega_{\nu}}\frac{d\Omega_{\nu}}{dE}& \simeq &
\left<\frac{{M_1^\star}}{{\Vert\nabla H\Vert}}\right>_{mc} =\, 
\frac{1}{\Omega_{\nu}} \int_{\Sigma_E} \frac{d\sigma}
{\Vert\nabla H\Vert}\, \frac{M_1^\star}{\Vert\nabla H\Vert}
= D(E)\frac{1}{\Omega_{\nu}} \int_{\Sigma_E} d\sigma \, M_1\nonumber \\
&\geq &  
\frac{D(E)}{\Omega_{\nu}}\left[ {\cal A}
\sum_{i=0}^{\nu} b_i(\Sigma_E ) + {\cal R}(E) \right]^{1/\nu}~~.
\label{thermtop}
\end{eqnarray} 
Equation (\ref{thermtop}) has the remarkable property of relating 
the microcanonical definition of
temperature of Eq.(\ref{deformaz}) with a {\it topologic invariant} of
$\Sigma_E$.
The Betti numbers can be exponentially large with $N$ [for example, 
in the case of $N$-tori ${\Bbb T}^N$, they are $b_k={N\choose k}$], so that 
the quantity $(\sum b_k)^{1/N}$ can converge, at arbitrarily large $N$, to a 
non-trivial limit (i.e. different from one). 
Thus, even though the energy dependence of ${\cal R}$ is
unknown, the energy variation of $\sum b_i(\Sigma_E)$ must be mirrored -- at 
any arbitrary $N$ -- by the energy variation of the temperature. 
By considering Eq.(\ref{cv2}) in the light of Eq.(\ref{thermtop}), we can
expect that 
some suitably abrupt and major change in the topology of the $\Sigma_E$ can 
reflect into the appearance of a peak of the specific heat,
as a consequence of the associated  
energy dependence of $\sum b_k(\Sigma_E)$ and of its derivative with respect 
to $E$. In other words, we see that a link must exist between thermodynamical
phase transitions and suitable topology changes of the constant energy 
submanifolds of the phase space of microscopic variables. 
The arguments given above, though in a still rough formulation, provide a 
first attempt to
make a connection between the topological aspects of the {\it microcanonical} 
description of phase transitions and the already proposed {\it topological 
hypothesis} about topology changes in configuration space and phase transitions
\cite{CCCP,CCCPPG,cegdcp,fps1,fps2}. 

Direct support to the {\it topological 
hypothesis} has been given by the analytic study of a mean-field XY model
\cite{cegdcp} and by the numerical computation of the Euler characteristic 
$\chi$ of the equipotential hypersurfaces $\Sigma_v$ of the configuration 
space in a $2d$ lattice $\varphi^4$ model \cite{fps2}.
The Euler characteristic is the alternate sum of all the Betti numbers of
a manifold, so it is another topological invariant, but it identically
vanishes for odd dimensional manifolds, like the $\Sigma_E$. 
In Ref.\cite{fps2},
$\chi (\Sigma_v)$ neatly reveals the symmetry-breaking phase transition
through a sudden change of
its variation rate with the potential energy density $v$. A sudden 
``second order'' variation of the topology of the $\Sigma_v$ appears in both
Refs.\cite{cegdcp,fps2} as the requisite for the appearance of a phase 
transition. These results strenghten the arguments given in the present 
Section about the role of the topology of the constant energy hypersurfaces.
In fact, the larger is $N$, the smaller are the relative fluctuations
$\langle\delta^2V\rangle^{1/2}/\langle V\rangle$ and $\langle\delta^2K\rangle
^{1/2}/\langle K\rangle$ of the potential and kinetic energies respectively.
At very large $N$,  the product manifold
$\Sigma^{N-1}_v\times{\Bbb S}_t^{N-1}$, with $v\equiv\langle V\rangle$ and
$t\equiv\langle K\rangle$, $v+t=E$, is a good model manifold to represent 
the part of $\Sigma_E$ that is overwhelmingly sampled by the dynamics and that
therefore constitutes the effective support of the microcanonical measure
on $\Sigma_E$. The kinetic energy submanifolds ${\Bbb S}_t^{N-1}=\{(p_1,\dots 
,p_N)\in{\Bbb R}^N\vert\sum_{i=1}^N\frac{1}{2}p_i^2=t\}$ are hyperspheres.

In other words, at very large $N$ the microcanonical measure 
mathematically extends over a whole energy surface but, as far as physics
is concerned, a non-negligible contribution to the microcanonical measure is
in practice given only by a small subset of an energy surface. This subset can
be reasonably modeled by the product manifold $\Sigma^{N-1}_v\times{\Bbb S}
_t^{N-1}$, because the total kinetic and total potential energies - having
arbitrarily small fluctuations, provided that $N$ is large enough - can be
considered almost constant. Thus, since ${\Bbb S}_t^{N-1}$ at any $t$ is always
an hypersphere, a change in the topology of $\Sigma^{N-1}_v$ directly entails 
a change of the topology of $\Sigma^{N-1}_v\times {\Bbb S}_t^{N-1}$, that is
of the effective model-manifold for the subset of $\Sigma_E$ where the dynamics
mainly ``lives'' at a given energy $E$.

At small $N$, the model with a single product manifold is no longer good and 
should be replaced by the non-countable union $\bigcup_{v\in{\cal I}\subset
{\Bbb R}}\Sigma^{N-1}_v\times{\Bbb S}_{E-v}^{N-1}$, with $v$ assuming all the 
possible values in a real interval ${\cal I}$. From this fact the smoothing 
of the energy dependence of thermodynamic variables follows. 
Nevertheless, the geometric and 
topologic signals of the phase transition can remain much sharper than the
thermodynamic signals also at small $N$ $(< 100)$, as it is witnessed by the
$2d$ lattice $\varphi^4$ model \cite{fps1,fps2}.

Finally, let us comment about the relationship between intrinsic geometry, 
in terms of
which we discussed the geometrization of the dynamics, and extrinsic
geometry, dealt with in the present Section.

The most direct and intriguing link is estabilished by the 
expression for microcanonical averages of generic observables of the kind
$A(q)$, with $q=(q_1,\dots,q_N)$,
\begin{equation}
\langle A\,\rangle_{mc}  = \frac{1}{\Omega_{2N}(E)} \int_{H(p,q)\leq E} 
d^Np\, d^Nq\, A(q)
=\frac{1}{Vol(M_E)}\int_{V(q)\leq E} d^Nq\, [E - V(q)]^{N/2}\, A(q)~,
\label{medie1}
\end{equation}
where $M_E=\{q\in{\Bbb R}^N\vert V(q)\leq E\}$. Eq. (\ref{medie1}) is 
obtained by means of a Laplace-transform method \cite{Pearson}; it is 
remarkable that $[E-V(q)]^N\equiv det(g_J)$, where $g_J$ is the Jacobi metric
whose geodesic flow coincides with newtonian dynamics (see Section $V$),
therefore $d^Nq\, [E - V(q)]^{N/2}\equiv d^N q\, \sqrt{det (g_J)}$
is the invariant Riemannian volume element of $(M_E,g_J)$. Thus, 
\begin{equation}
\frac{1}{Vol(M_E)}\int_{V(q)\leq E} d^Nq\, [E - V(q)]^{N/2}\, A(q) \equiv\;
\frac{1}{Vol(M_E)}\int_{M_E} d^Nq\, \sqrt{det (g_J)}\, A(q)~,
\label{medie2}
\end{equation} 
which means that the microcanonical averages $\langle A(q)\,\rangle_{mc}$ 
can be expressed as Riemannian integrals on the mechanical manifold 
$(M_E,g_J)$. 

In particular, this also applies to the microcanonical definition of entropy 
\begin{equation}
S = k_B \log \int_{H(p,q)\leq E} d^N p\, d^N q\, =
 k_B \log \int_0^E dE^\prime \int_{\Sigma_{E^\prime}} \frac{d\sigma}
{\Vert{\nabla H}\Vert}~,  
\label{entropy1}
\end{equation}
which is alternative to that given in Eq.(\ref{entropy}), though equivalent 
to it in the large $N$ limit. We have  
\begin{eqnarray}
S &=& k_B \log \left[ \frac{1}{C \Gamma (N/2+1)} \int_{V(q)\leq E} 
d^Nq \;[E - V(q)]^{N/2}\right] \nonumber \\
&\equiv & k_B \log \int_{M_E} d^N q\, \sqrt{det (g_J)}+\, const\,.~~,
\label{entropy2}
\end{eqnarray}
where the last term gives the entropy as the logarithm of the 
Riemannian volume  of the manifold.  

The topology changes of the surfaces $\Sigma_v^{N-1}$, that are to be 
associated with phase transitions, will deeply affect also the geometry of
the mechanical manifolds $(M_E, g_J)$ and $(M\times{\Bbb R}^2, g_E)$ 
and, consequently, they will affect the average instability properties of
their geodesic flows. In fact, Eq.(\ref{lambda}) links some curvature
averages of these manifolds with the numeric value of the largest Lyapunov 
exponent. Loosely speaking, major topology changes of $\Sigma_v^{N-1}$ will
affect microcanonical averages of geometric quantities 
computed through Eq.(\ref{medie1}), likewise entropy, computed
through Eq.(\ref{entropy2}).

Thus, the peculiar temperature patterns displayed by the largest Lyapunov 
exponent at a second-order phase transition point -- in the present paper
reported for the $3d$ $XY$ model, in Ref.\cite{CCCPPG} reported for
lattice $\varphi^4$ models -- appear as reasonable consequences of the deep 
variations of the topology of the equipotential hypersurfaces of configuration 
space.

We notice that topology seems to provide a common ground to the roots of 
microscopic dynamics and of thermodynamics and, notably, it can account for
 major qualitative changes simultaneously occurring in both dynamics and
thermodynamics when a phase transition is present.

\medskip
\subsection{Some preliminary numerical computations}
\label{5a}
\medskip
Let us briefly report on some preliminary numerical computations concerning 
the extrinsic geometry of the hypersurfaces $\Sigma_E$ in the case of the $3d$ 
XY model. 

The first point about extrinsic geometry that we numerically addressed was to
check whether the inverse of the temperature, that appears in 
Eq.(\ref{deformaz}), can be reasonably factorized into the product of a smooth 
``deformation factor'' $D(E)$ and of the total mean curvature 
$\int_{\Sigma_E}M_1d\sigma$.
To this purpose, the two independently computed quantities 
$\langle 1/\Vert\nabla H\Vert^2\rangle_{mc}$ and $D(E)=[\int_{\Sigma_E}
(d\sigma /\Vert\nabla H\Vert )(M_1^\star/\Vert\nabla H\Vert)]\,/\,
[\int_{\Sigma_E}d\sigma M_1]$ are compared in Fig. \ref{D(E)},
showing that actually $\int_{\Sigma_E}(d\sigma /\Vert\nabla H\Vert )
(M_1^\star/\Vert\nabla H\Vert) \simeq
\langle 1/\Vert\nabla H\Vert^2\rangle_{mc}\int_{\Sigma_E}d\sigma M_1$.
In other words, $D(E)\simeq\langle 1/\Vert\nabla H\Vert^2\rangle_{mc}$ 
and no ``singular'' feature in its energy pattern
seems to exist, what suggests that $\int_{\Sigma_E} d\sigma \ M_1$ has to 
convey all the information 
relevant to the detection of the phase transition.
There is no reason to think that the validity of the 
factorization given in Eq.(\ref{deformaz}) is limited to the special case 
of the XY model.

The other point that we tackled concerns an indirect quantification of how
a phase space trajectory curves around and knots on the $\Sigma_E$ to which 
it belongs. We can expect that the way in which an hypersurface 
$\Sigma_E$ is ``filled'' by a phase space trajectory living on it will be 
affected by the geometry and the topology of the $\Sigma_E$.
In particular, we computed the normalized autocorrelation function of the time
series $M_1[x(t)]$ of the mean curvature at the points of $\Sigma_E$ visited
by the phase space trajectory, 
that is, the quantity
\begin{equation}
\Gamma (\tau )=\langle \delta M_1(t+\tau )\delta M_1(t)\rangle_t~~,
\label{autocor}
\end{equation}
where $\delta M_1(t)=M_1(t)-\langle M_1(t^\prime )\rangle_{t^\prime}$ is the
fluctuation with respect to the average (the ``process'' $M_1(t)$ is supposed
stationary). Our aim was to highlight the extrinsic
geometric-dynamical counterpart of a symmetry-breaking phase transition.

The practical computation of $\Gamma (\tau )$ proceeds by working out the 
Fourier power spectrum $\vert \tilde M_1(\omega )\vert^2$ of $M_1[x(t)]$,
obtained by averaging $15$ spectra computed by an FFT algorithm with a mesh
of $2^{15}$ points and a sampling time $\Delta t=0.1$.
Some typical results for $\Gamma (\tau )$, obtained at different temperatures,
are reported in Fig.\ref{Gamma}. The patterns $\Gamma (\tau )$ display a
first regime of very fast decay, which is not surprising because of the 
chaoticity of the trajectories at 
any energy, followed by a longer tail of slower 
decay. An autocorrelation time $\tau_{corr}$ can be defined
through the first intercept of $\Gamma (\tau )$ with an almost-zero level
($\Gamma =0.01$).
In Fig.\ref{tau} we report the values of $\tau_{corr}$ so defined vs.
temperature.
In correspondence of the phase transition (whose critical temperature is 
marked by a vertical dotted line), $\tau_{corr}$ changes its temperature 
dependence: by lowering the temperature,
below the transition $\tau_{corr}(T)$ rapidly increases, 
whereas it mildly 
decreases above the transition. Below $T\simeq 0.9$, where the vortices
disappear, the autocorrelation functions of $M_1$ look quite different and 
it seems no longer possible to coherently define a correlation time. 
This result has an intuitive meaning and confirms that the phase transition
corresponds to a change in the microscopic dynamics, as already signaled by
the largest Lyapunov exponent; however, notice that the correlation times 
$\tau_{corr}(T)$ are much longer than the inverse values of the corresponding 
$\lambda_1(T)$. Qualitatively, $\lambda_1(T)$ and $\tau_{corr}^{-1}(T)$
look similar, however the two functions are not simply related.
\medskip
\section{Discussion and perspectives}
\medskip
The microscopic Hamiltonian dynamics of the classical Heisenberg XY model in
two and three spatial dimensions has been numerically investigated.
This has been possible after the addition to the Heisenberg potentials of
a standard (quadratic) kinetic energy term.
Special emphasis has been given to the study of the dynamical counterpart
of phase transitions, detected through the time averages of conventional 
thermodynamic observables, and to the new mathematical concepts that are
brought about by Hamiltonian dynamics.

The motivations of the present study are given in the Introduction.
Let us now summarize what are the outcomes of our investigations and comment 
about their meaning.  There are three main topics, tightly related one to 
the other:
\begin{itemize} 
\item{} the phenomenological description of phase transitions through the 
natural, microscopic dynamics in place of the usual Monte Carlo stochastic 
dynamics; 
\smallskip
\item{} the investigation, in presence of phase transitions, of certain  
aspects of the (intrinsic) geometry of the mechanical manifolds where the
natural dynamics is represented as a geodesic flow;
\smallskip
\item{} the discussion of the relationship between the (extrinsic) 
geometry of constant energy hypersurfaces of phase space and thermodynamics.
\smallskip
\end{itemize}
About the first point, we have found that microscopic Hamiltonian dynamics very
clearly evidences the presence of a second order phase transition through the
time averages of conventional thermodynamic observables. Moreover, the 
familiar 
sharpening effects, at increasing $N$, of the specific heat peak and  
of the order parameter bifurcation are observed. The evolution of the 
order parameter with respect to the physical time (instead of the
fictitious Monte Carlo time) is also accessible, showing the appearance of 
Goldstone modes and that, in presence of a second order phase transition,
there is a clear tendency to the freezing of transverse fluctuations
of the order parameter when $N$ is increased. 
The "freezing" is observed together with a reduction of the
longitudinal fluctuations, i.e.  the rotation of the magnetization 
vector slows down, preparing the breaking of the $O(2)$ symmetry at 
$N\rightarrow\infty$. At variance, when a Kosterlitz-Thouless transition is 
present, at increasing $N$ the magnetization vector has a faster rotation
and a smaller norm, preparing the absence of symmetry-breaking in the
$N\rightarrow\infty$ limit as expected. 

Remarkably, to detect phase transitions, microscopic Hamiltonian dynamics 
provides us with additional observables of purely dynamical nature, i.e.
without statistical counterpart: Lyapunov exponents. Similarly to what we and 
other authors already reported for other models (see Introduction), also
in the case of the $3d$ XY model a peculiar 
temperature pattern of the largest Lyapunov exponent shows up in presence of
the second order phase transition, signaled by a ``cuspy'' point. 
By comparing the patterns $\lambda_1(T)$ given by Hamiltonian dynamics and by 
a suitably defined random dynamics respectively, we suggest that the 
transition between thermodynamically ordered and disordered phases has its 
microscopic dynamical counterpart in a transition between weak and strong 
chaos. Though $a-posteriori$ physically reasonable, this result is far from
obvious, because the largest Lyapunov exponent measures the average 
{\it local instability} of the dynamics, which $a-priori$ has little to do
with a {\it collective}, and therefore global, phenomenon such as a phase 
transition.
The effort to understand the reason of such a sensitivity of $\lambda_1$ to
a second order phase transition and to other kinds of transitions, as
mentioned in the Introduction, is far reaching. 

Here we arrive to the second
point listed above. In the framework of a Riemannian geometrization of
Hamiltonian dynamics, the largest Lyapunov exponent is related to the 
curvature properties of suitable submanifolds of configuration space whose
geodesics coincide with the natural motions. In the mathematical light of this
geometrization of the dynamics, and after the numerical evidence of a sharp
peak of curvature fluctuations at the phase transition point, the peculiar 
pattern of $\lambda_1(T)$ is due to some major change occurring to the geometry
of mechanical manifolds at the phase transition. Elsewhere, we have conjectured
that indeed some major change in the {\it topology} of configuration space
submanifolds should be the very source of the mentioned major change of 
geometry. 

Thus, we have made a first attempt to provide an analytic argument
supporting this topological hypothesis (third point of the above list).
This is based on the appearance of a non trivial relationship between the
geometry of constant energy hypersurfaces of phase space with their topology
and with the microcanonical definition of thermodynamics. Even still in a
preliminary formulation, our reasoning already seems to indicate  the 
topology of energy hypersurfaces as the best candidate to explain the deep
origin of the dynamical signature of phase transitions detected through
$\lambda_1(T)$.

The circumstance, mentioned in the preceding Section, of the persistence 
at small $N$ of  geometric and topologic signals of the phase transition 
that are much sharper than the thermodynamic signals is of
prospective interest for the study of phase transition phenomena in finite,
small systems, a topic of growing interest thanks to the modern developments
- mainly experimental - about the physics of nuclear, atomic and molecular 
clusters, of conformational phase transitions in homopolymers and proteins,
of mesoscopic systems, of soft-matter systems of biological interest.
In fact, some unambiguous information for small systems - even about the 
existence itself of a phase transition - could be better obtained by means 
of concepts and mathematical tools outlined here and in the quoted papers.
Here we also join the very interesting line of thought of Gross and 
collaborators \cite{Gross,Gross3} about the microcanonical description of phase
transitions in finite systems. 

Let us conclude with a speculative comment about another possible direction of
investigation related with this signature of phase transitions through Lyapunov
exponents. 
In a field-theoretic framework, based on a path-integral
formulation of classical mechanics \cite{reuter,gozzi1,gozzi2}, Lyapunov 
exponents are defined through the expectation values of suitable operators. 
In the field-theoretic framework, ergodicity breaking appears related to a
supersymmetry breaking \cite{reuter}, and Lyapunov exponents are related to 
mathematical objects that have many analogies with topological concepts 
\cite{gozzi2}.

The new mathematical concepts and methods, that the Hamiltonian 
dynamical approach brings about, could hopefully be useful also in the study 
of more ``exotic'' transition phenomena than those tackled in the present 
work. Besides the above mentioned soft-matter systems, this could be the case 
of transition phenomena occurring in amorphous and disordered materials.

\medskip
\section{acknowledgments}
\medskip
We warmly thank L. Casetti, E.G.D. Cohen,R. Franzosi and L. Spinelli for
many helpful discussions.
During the last year C.C. has been supported by the
NSF (Grant \# 96-03839) and by the La Jolla Interfaces in Science
program (sponsored by the Burroughs Wellcome Fund).
This work has been partially supported by I.N.F.M., under the PAIS
{\it Equilibrium and non-equilibrium dynamics in condensed matter systems},
which is hereby gratefully acknowledged.

\medskip
\section{Appendix A}
\medskip

Let us briefly explain how a random markovian dynamics is constructed on a
given constant energy hypersurface of phase space. The goal is to
compare the energy dependence of the largest Lyapunov exponent computed
for the Hamiltonian flow and for a suitable random walk respectively.
One has to devise an algorithm to generate a random walk on a given energy
hypersurface such that, once the time interval $\Delta t$ separating two
successive steps is assigned, the average increments of the coordinates
are equal to the average increments of the same coordinates for the
differentiable dynamics integrated with a time step $\Delta t$.
In other words, the random walk has to roughly mimick the differentiable 
dynamics with the exception of its possible time-correlations.

One starts with a random initial configuration of the coordinates
$q_{i},~ i=1,2,\ldots,N$, uniformly distributed in the interval
$[0,2\pi]$, and with a random gaussian-distributed choice of the 
coordinates $p_{i}$.
The random pseudo-trajectory is generated according to the simple scheme
\begin{eqnarray}
(q_{i})_{(k+1)\Delta t} &\mapsto& (q_{i})_{k \Delta t} +
\alpha _{q} G_{i,k} \Delta t \nonumber\\
(p_{i})_{(k+1)\Delta t} &\mapsto& (p_{i})_{k \Delta t} +
\alpha _{p} G_{i,k} \Delta t~~,
\label{step.micro}
\end{eqnarray}
where $\Delta t$ is the time interval associated to one step $k \mapsto k+1$ 
in the markovian chain, $G_{i,k}$ are gaussian distributed random numbers with
zero expectation value and unit variance; the parameters
$\alpha_{q}$ and $\alpha_{p}$ are the variances of the processes
$(q_i)_k$ and $(p_i)_k$.
These variances are functions of the energy per degree of freedom
$\varepsilon$. They have to be set equal to the 
numerically computed average increments of the coordinates obtained  
along the differentiable trajectories integrated with the same time step 
$\Delta t$, that is
\begin{eqnarray}
\alpha_{q}(\epsilon)&=&\left\langle\left[{\frac{1}{N}\sum_{i=1}^{N}
\frac{(q_{i}(t+\Delta t)
-q_{i}(t))^{2}}{\Delta t}}\right]^{1/2}\right\rangle_t \sim
\left\langle\left[ {\frac{1}{N}\sum_{i=1}^{N}p_{i}^{2}}\right]^{1/2}
\right\rangle_t\sim \sqrt{T}\nonumber\\
\alpha_{p}(\epsilon)&=&\left\langle\left[{\frac{1}{N}\sum_{i=1}^{N}
\frac{(p_{i}(t+\Delta t)-p_{i}(t))^{2}}{\Delta t}}\right]^{1/2}\right
\rangle_t\sim
\left\langle\left[{\frac{1}{N}\sum_{i=1}^{N} \dot{p}_{i}^{2}}\right]^{1/2}
\right\rangle_t~~,
\end{eqnarray}
where $T$ is the temperature.
Then, in order to make minimum the energy fluctuations around any given value
of the total energy, a criterium to accept or reject a new step along the
markovian chain has to be assigned.
A similar problem has been considered by Creutz, who developed
a Monte Carlo microcanonical algorithm \cite{Creutz}, where a 
"Maxwellian demon" gives a part of its energy to the system to let it move to 
a new configuration, or gains energy from the system,
if the new proposed configuration produces an energy lowering. If the demon
does not have enough energy to allow an energy increasing update of the
coordinates, no coordinate change is performed. In this way, the total 
energy remains almost constant with only small fluctuations.
As usual in Monte Carlo simulations, it is appropriate to fix the parameters 
so as the acceptance rate of the proposed updates of the configurations is 
in the range $30\%$ -- $60\%$.

A reliability check of the so defined random walk, and of the 
adequacy of the phase space sampling through the number of steps 
adopted in each run, is obtained by computing the averages of typical 
thermodynamic observables of known temperature dependences.

An improvement to the above described ``demon'' algorithm  has been 
obtained through a simple reprojection on $\Sigma_E$ of the
updated configurations \cite{Pettini}; 
the coordinates generated by means of (\ref {step.micro})
are corrected with the formulae
\beq
q_{i}(k \Delta t) \mapsto q_{i}(k \Delta t) +
\left[\frac{(\frac{\partial H}{\partial q_{i}}) \Delta E}
{\sum_{i=1}^N (p_{j}^2+
(\frac{\partial H}{\partial q_{j}})^2 )}\right]_{x_{R}({k \Delta t})} \
\eeq
\[
p_{i}({k \Delta t}) \mapsto p_{i}({k \Delta t}) -
\left[\frac{p_{i} \Delta E}{\sum_{j=1}^N (p_{j}^2+
(\frac{\partial H}{\partial q_{j}})^2 )}\right]_{x_{R}({k \Delta t})}~~, \
\]
where $\Delta E$ is the difference between the energy of the new configuration
and the reference energy, and $x_{R}({k \Delta t})$ denotes the random
phase space trajectory.
At each assigned energy, the computation of the largest Lyapunov exponent
$\lambda_{1}^R$ of this random trajectory is obtained by means of the 
standard definition
\begin{equation}
\lambda_{1}^R = \lim_{n \rightarrow \infty} \frac{1}{n \Delta t}\sum_{k=1}^n
\log \frac{\| \zeta ((k+1) \Delta t)\|}{\| \zeta (k \Delta t)\|}~~,
\label{bgs}
\end{equation}
where $\zeta (t)\equiv (\xi (t),\dot\xi (t))$ is given by the discretized 
version of the tangent dynamics
\begin{equation}
\frac{\xi_i ((k+1)\Delta t)-2 \xi_i(k\Delta t)+\xi_i ((k-1)\Delta t)}
{\Delta {t}^2}+
\left(\frac{\partial^2 V}{\partial q_{i} \partial q_{j}}\right)_
{x_{R}(k \Delta t)}
\xi_j (k \Delta t) = 0~~.
\label{tandynR}
\end{equation}

For wide variations of the parameters ($\Delta t$ and
acceptance rate), 
the resulting values of $\lambda_{1}^R$ are in very good agreement.
Moreover, the algorithm is sufficiently stable
 and the final value of $\lambda_{1}^R$ is independent of
the choice of the initial condition.

A more refined algorithm could be implemented by constructing a random 
markovian process $q(t_k)\equiv [q_1(t_k),\dots ,q_N(t_k)]$ performing an 
importance sampling of the measure $d\mu =[E-V(q)]^{N/2-1}\,dq$ in 
configuration space. In fact, similarly to what is reported in 
Eq.(\ref{medie1}), one has \cite{Pearson} $\int_{H(p,q)= E} d^Np\, d^Nq\, =
\, const\,\int_{V(q)\leq E} d^Nq\, [E - V(q)]^{N/2-1}$. A random  process
obtained by sampling such a measure -- with the additional property
of a relation between the average increment and the physical time step 
$\Delta t$ as discussed above --
 would enter into Eq.(\ref{tandynR}) to yield $\lambda_1^R$.
However, this would result in much heavier numerical computations (with some
additional technical difficulty at large $N$) which was not worth in view of 
the principal aims of the present work.

\newpage

\clearpage

\clearpage

\clearpage

\clearpage

\clearpage

\clearpage

\clearpage

\clearpage

\clearpage

\clearpage

\clearpage

\clearpage

\clearpage

\begin{figure}
\centerline{\includegraphics[width=0.80\linewidth]{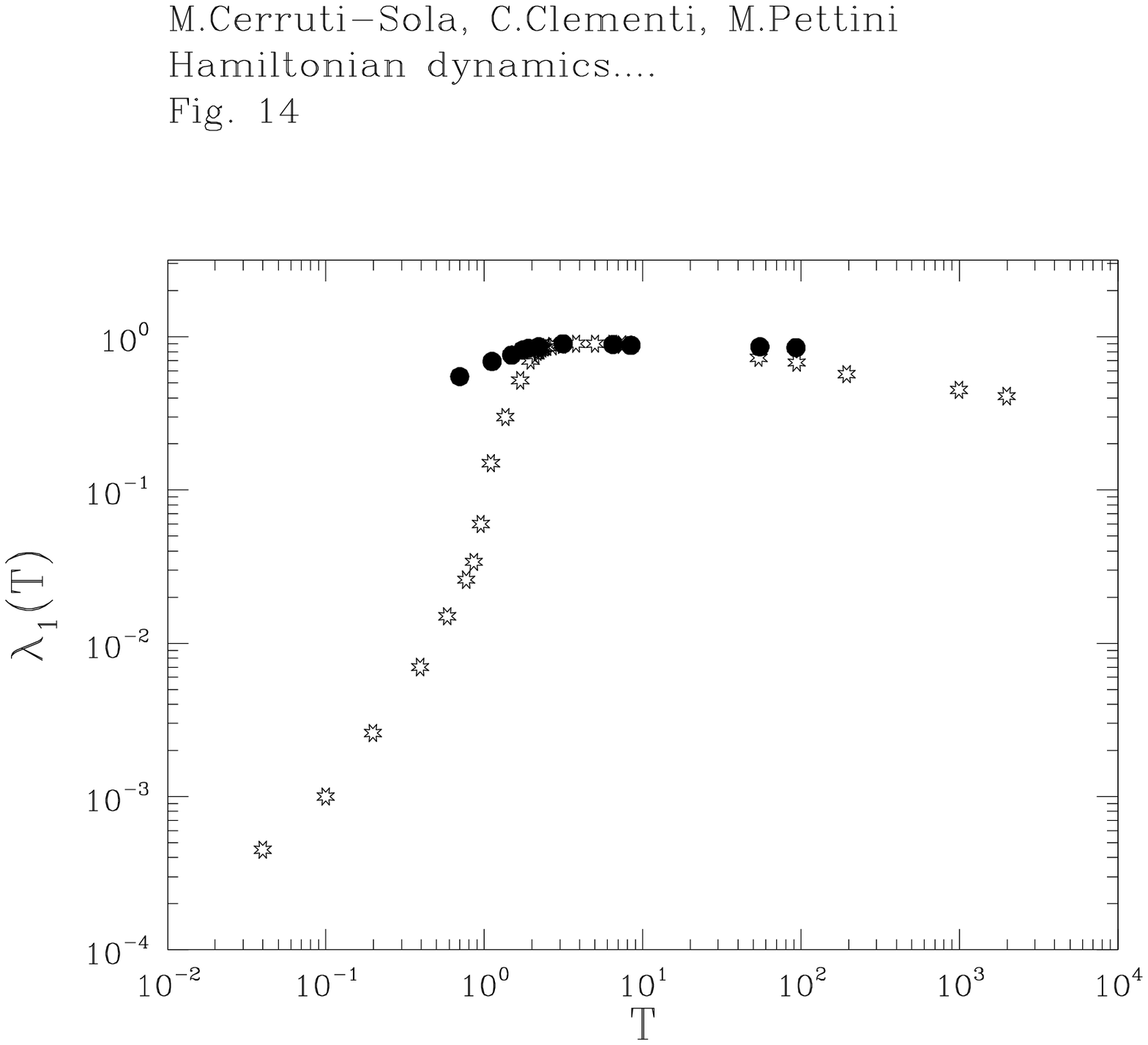}}
\caption{ The largest Lyapunov exponents computed by means of the random 
dynamics algorithm (full circles) are plotted 
 in comparison with those computed by means of the standard dynamics
(open stars) for a lattice of $N= 10 \times 10 \times 10$. }
\label{randyn.3d.fig}
\end{figure}
\clearpage

\begin{figure}
\centerline{\includegraphics[width=0.80\linewidth,angle=90]{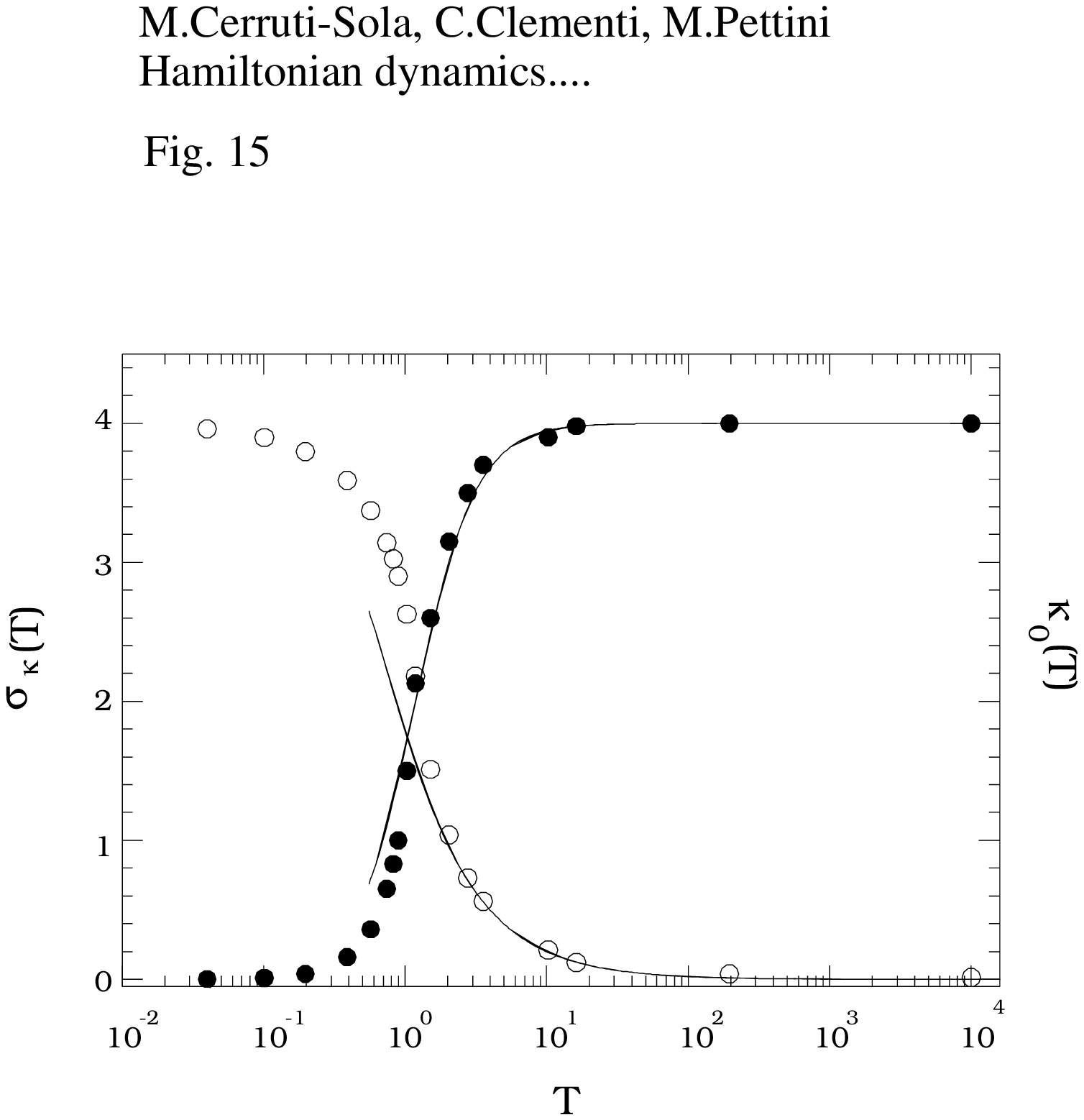}}
\caption{ Time average of Ricci curvature (open circles) and its r.m.s.
 fluctuations
(full circles) at different temperatures computed for a lattice
of $N = 40 \times 40$ sites. Solid lines are the analytic estimates obtained 
from a high temperature expansion. } 
\label{fig.Ricci2d}
\end{figure}
\clearpage

\begin{figure}
\centerline{\includegraphics[width=0.80\linewidth,angle=90]{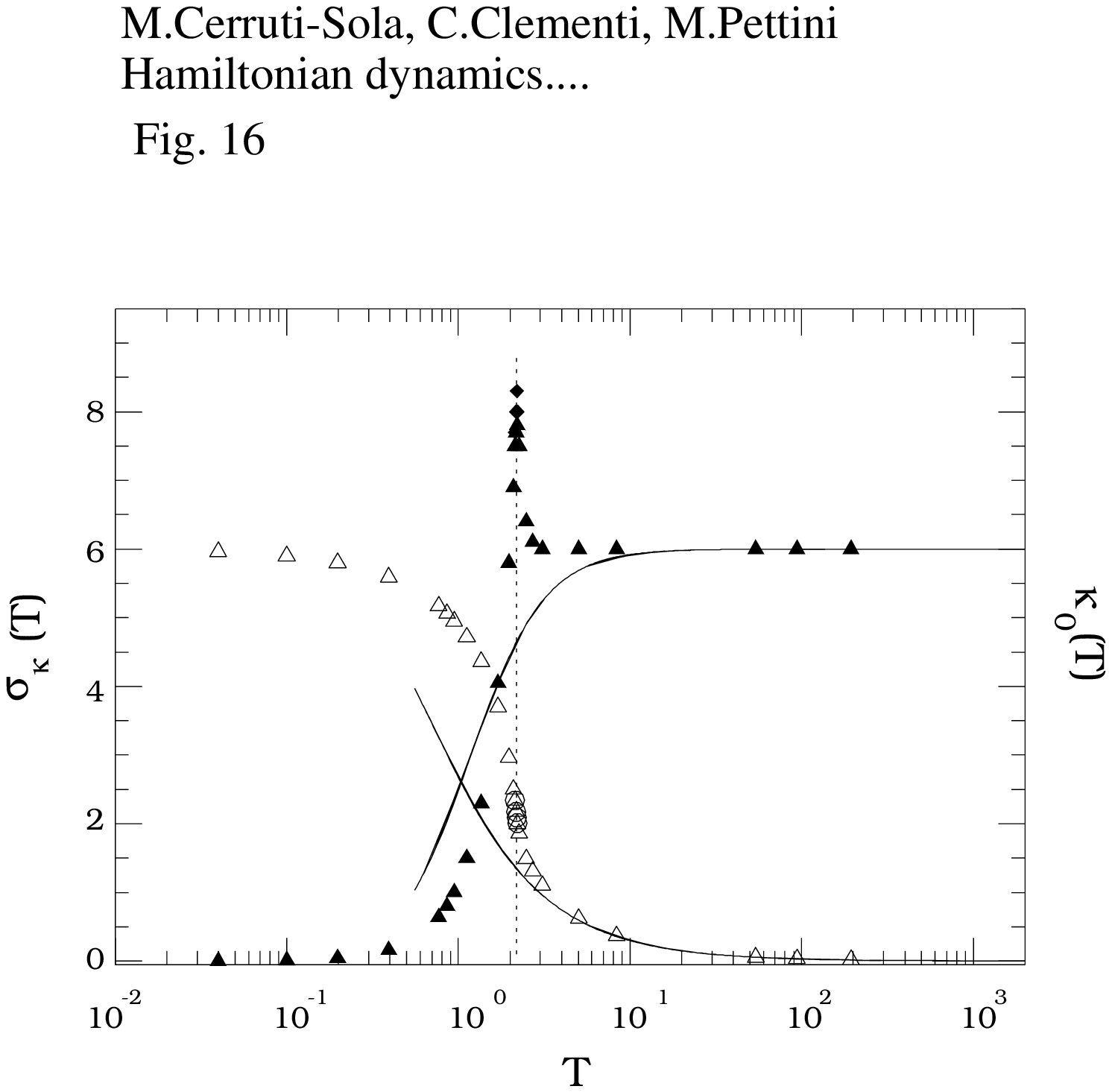}}
\caption{ Time average of Ricci curvature (open triangles) and its r.m.s.
 fluctuations
(full triangles) computed at different
temperatures for a lattice of $N = 10 \times 10 \times 10$.
Open circles and full diamonds refer to a lattice size 
of $N= 15 \times 15 \times 15$. Solid lines are the analytic estimates
in the limit of high temperatures. The dashed line points out the temperature
$T_c \simeq 2.17$ of the phase transition. }
\label{keflutt.3dfig}
\end{figure}
\clearpage

\begin{figure}
\centerline{\includegraphics[width=0.80\linewidth]{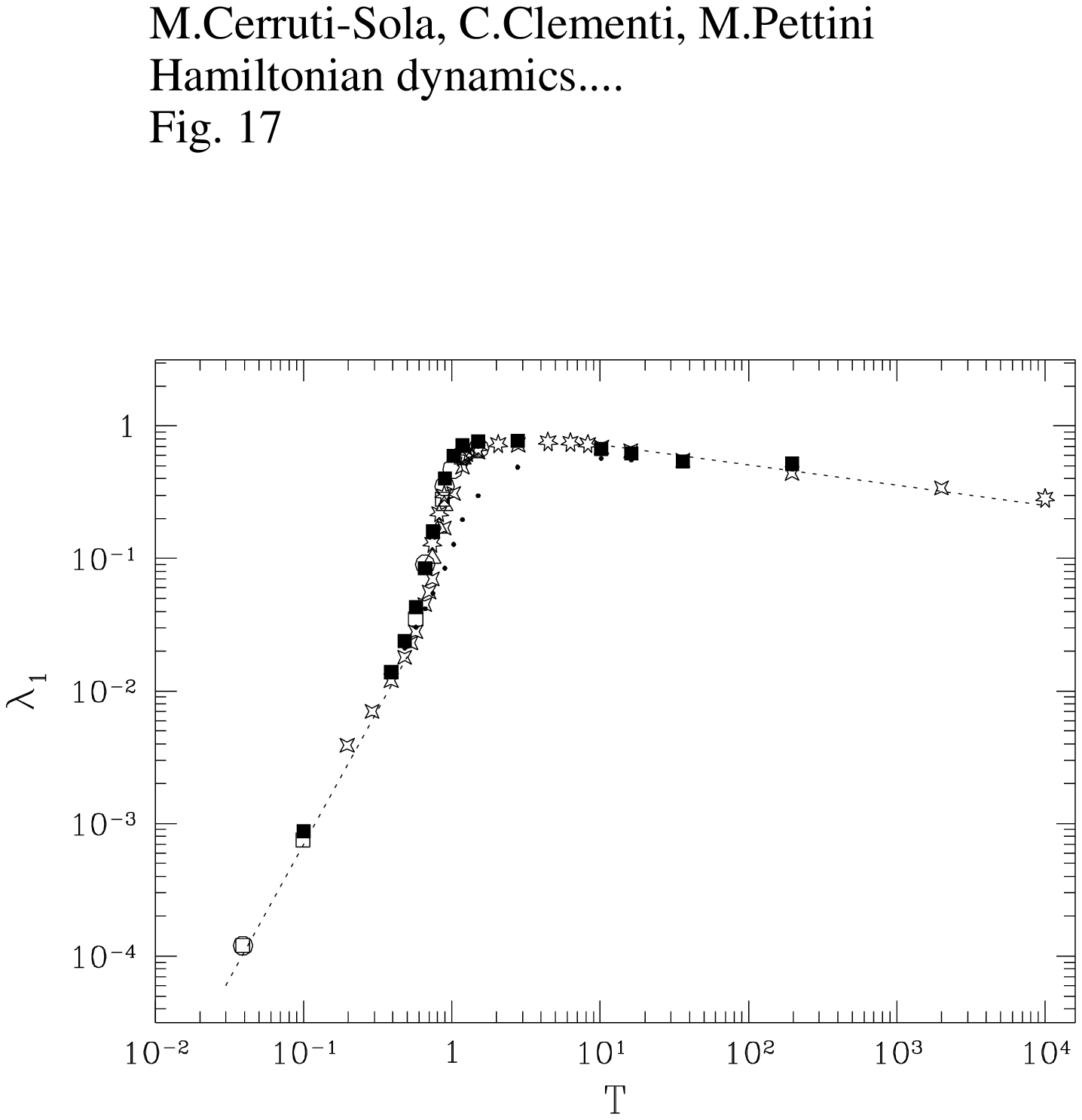}}
\caption{ Analytic Lyapunov exponents  computed for the $2d$ model by means of 
Eq.(\protect\ref{lambda}) without correction (dots) and incorporating  
the correction that accounts for
the probability of obtaining negative sectional curvatures
(full squares) for a lattice size of
$N = 40 \times 40$
are plotted in comparison with the numerical values 
of Fig. \ref{xy2d.lyap.num.fig}.
 The dashed lines are the asymptotic behaviors at high and low
temperatures in the thermodynamic limit. }
\label{prev.Lyap.2d}
\end{figure}
\clearpage

\begin{figure}
\centerline{\includegraphics[width=0.80\linewidth]{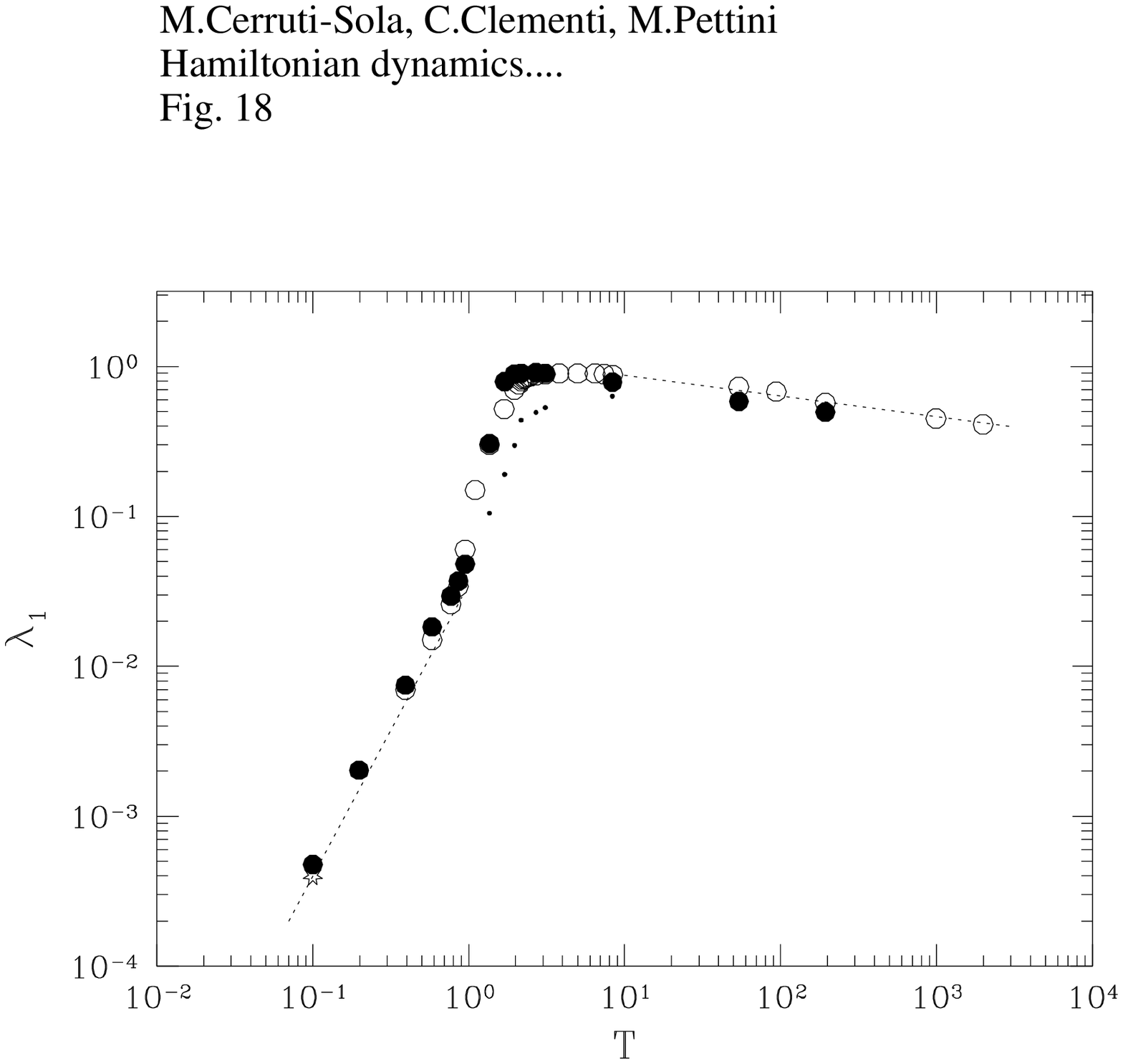}}
\caption{ Analytic Lyapunov exponents computed for the $3d$ model by means of 
Eq.(\protect\ref{lambda}) without correction (dots) and
incorporating
the correction that accounts for
the probability of obtaining negative sectional curvatures (full circles)
are plotted in comparison with the numerical values of
Fig. \ref{lyap.3d.fig}.
The dashed lines are the
asymptotic behaviors at high and low temperatures in the thermodynamic 
limit. }
\label{prev.Lyap.3d}
\end{figure}
\clearpage

\begin{figure}
\centerline{\includegraphics[width=0.80\linewidth]{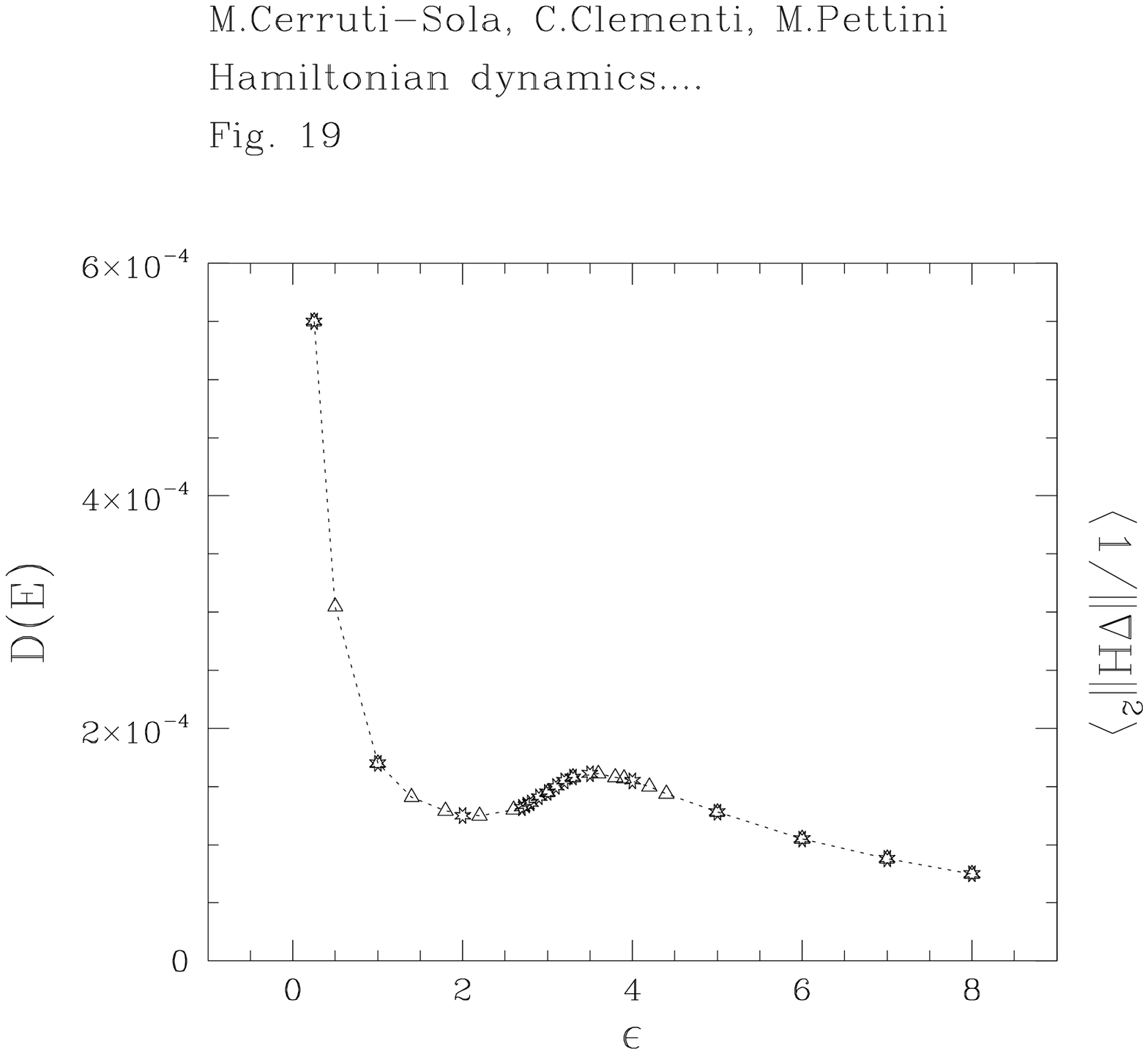}}
\caption{ The deformation factor $D(E)=[\int_{\Sigma_E}(d\sigma /\Vert
\nabla H\Vert )(M_1^\star/\Vert\nabla H\Vert)]\,/\,
[\int_{\Sigma_E}d\sigma M_1]$ of Eq. (\ref{deformaz}) (open circles)
is plotted vs. energy density $E/N$ and compared to the quantity
$\langle 1/\Vert\nabla H\Vert^2\rangle$ 
(open triangles). $N=10 \times 10\times 10$. }
\label{D(E)}
\end{figure}
\clearpage

\begin{figure}
\centerline{\includegraphics[width=0.80\linewidth]{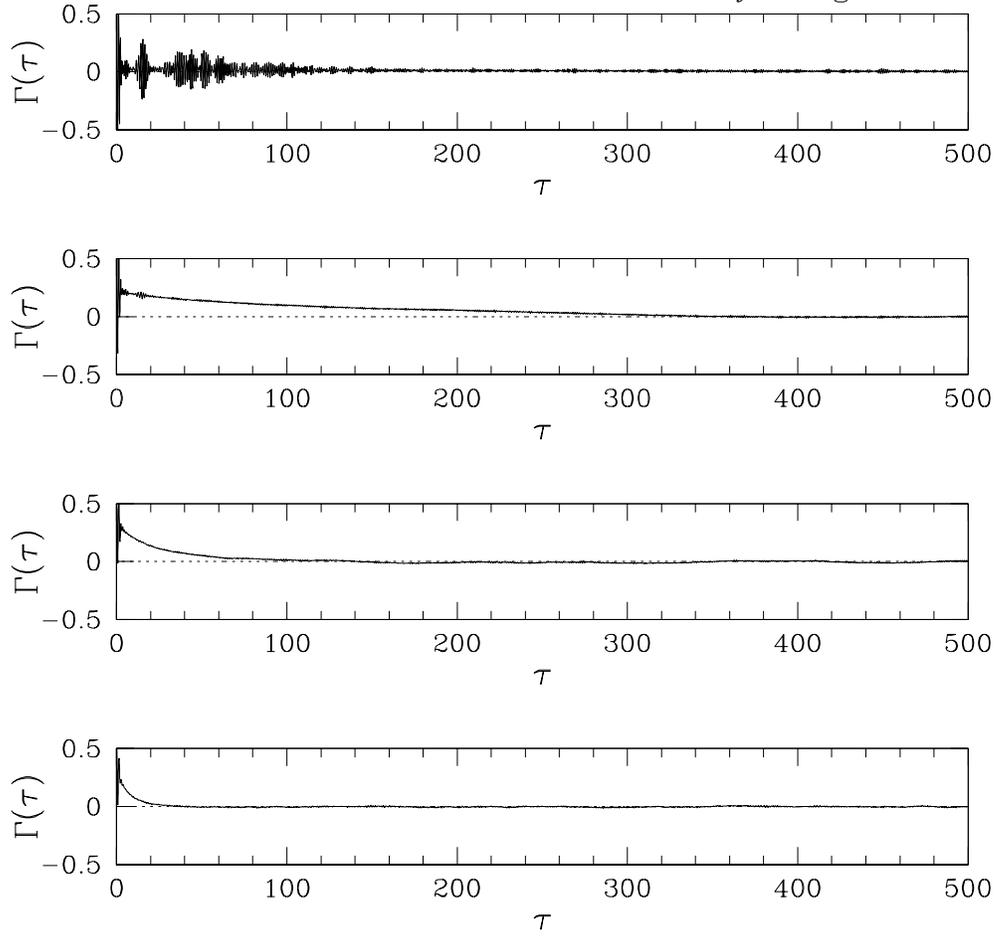}}
\caption{ The normalized autocorrelation functions $\Gamma(\tau)$ are plotted
vs. time $\tau$ for a lattice of $N=10 \times 10\times 10$ and for 
four different 
values of the temperature (from top to bottom: $T=0.49, 1.28, 1.75, 2.16$). }
\label{Gamma}
\end{figure}
\clearpage

\begin{figure}
\centerline{\includegraphics[width=0.80\linewidth]{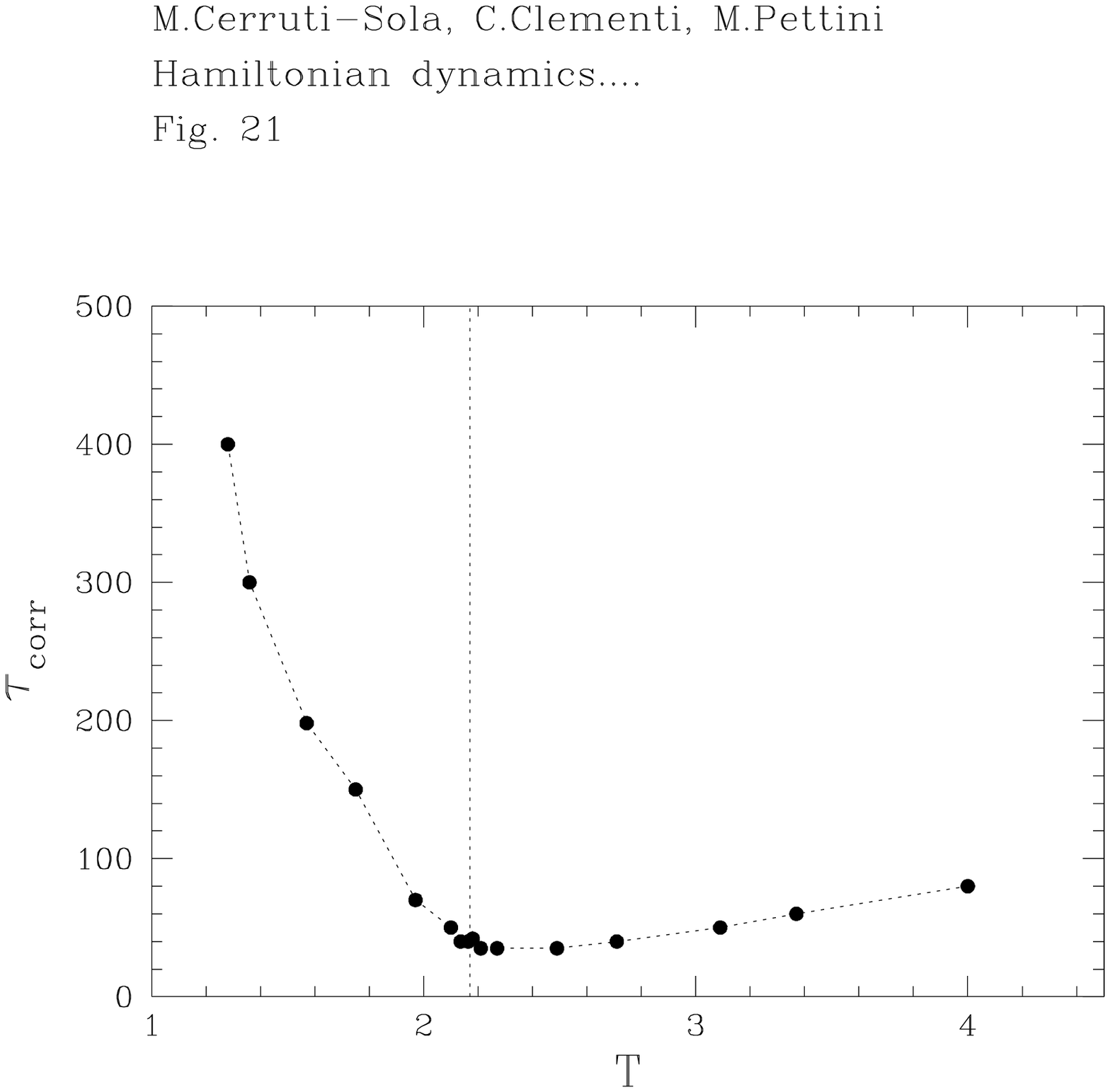}}
\caption{ Autocorrelation times $\tau_{corr}$ are plotted vs. temperature $T$.
The vertical dashed line points out the temperature $T_c \simeq2.17$ at
which the phase transition occurs. }
\label{tau}
\end{figure}
\clearpage
\end{document}